\documentclass[sigconf]{acmart}
\AtBeginDocument{%
  }

\setcopyright{acmlicensed}
\copyrightyear{2018}
\acmYear{2018}
\acmDOI{XXXXXXX.XXXXXXX}
\acmConference[Conference acronym 'XX]{Make sure to enter the correct
  conference title from your rights confirmation email}{June 03--05,
  2018}{Woodstock, NY}
\acmISBN{978-1-4503-XXXX-X/2018/06}

\usepackage[table]{xcolor}
\usepackage{multirow}
\usepackage{graphicx}  
\usepackage{subcaption}

\usepackage{amsthm}
\newtheorem{proposition}{Proposition}

\newtheorem{remark}{Remark}

\begin{document}

\title{Dual-Perspective Disentangled Multi-Intent Alignment for Enhanced Collaborative Filtering}

\author{Shanfan Zhang}
\affiliation{%
  \institution{School of software Engineering, \ Xi'an Jiaotong University}
  \city{Xi'an}
  \state{Shaanxi}
  \country{China}
}
\email{zhangsfxajtu@gmail.com}

\author{Yongyi Lin}
\affiliation{%
  \institution{School of Mathematics and Statistics, \ Xi’an Jiaotong University}
  \city{Xi'an}
  \country{China}}
\email{linyongyi@stu.xjtu.edu.cn}

\author{Yuan Rao}
\authornote{Corresponding author.}
\affiliation{%
  \institution{School of software Engineering, \ Xi'an Jiaotong University}
  \city{Xi'an}
  \state{Shaanxi}
  \country{China}
}
\email{raoyuan@mail.xjtu.edu.cn}

\author{Bingcan Xia}
\affiliation{%
  \institution{School of software Engineering, \ Xi'an Jiaotong University}
  \city{Xi'an}
  \state{Shaanxi}
  \country{China}
}
\email{bingcan92@stu.xjtu.edu.cn}

\author{Tingting Xin}
\affiliation{%
  \institution{School of software Engineering, \ Xi'an Jiaotong University}
  \city{Xi'an}
  \state{Shaanxi}
  \country{China}
}
\email{Tingtingxin@stu.xjtu.edu.cn}

\author{Chenlong Zhang}
\affiliation{%
  \institution{School of software Engineering, \ Xi'an Jiaotong University}
  \city{Xi'an}
  \state{Shaanxi}
  \country{China}
}
\email{zcl2023@stu.xjtu.edu.cn}

\renewcommand{\shortauthors}{Shanfan Zhang et al.}

\begin{abstract}
  
  Personalized recommendation requires capturing the complex latent intents underlying user–item interactions. Existing structural models, however, often fail to preserve perspective-dependent interaction semantics and provide only indirect supervision for aligning user and item intents, lacking explicit interaction-level constraints. This entangles heterogeneous interaction signals, leading to semantic ambiguity, reduced robustness under sparse interactions, and limited interpretability. To address these issues, we propose \emph{DMICF}, a Dual-Perspective Disentangled Multi-Intent framework for collaborative filtering. \emph{DMICF} models interactions from complementary user- and item-centric perspectives and employs a macro–micro prototype-aware variational encoder to disentangle fine-grained latent intents. Interaction-level supervision enforces dimension-wise alignment between user and item intents, grounding latent factors and enabling their collaborative emergence. Importantly, each component is architecturally flexible, and performance is robust to specific module instantiations. We offer a theoretical analysis to help explain how prototype-aware conditioning may alleviate posterior collapse, while the reconstruction objective promotes intent-wise contrastive alignment between positive and negative interactions. Extensive experiments on multiple benchmarks demonstrate consistent improvements over strong baselines, with ablations validating each core component. All code and datasets used are publicly available at https://anonymous.4open.science/r/code-5AA1/.

\end{abstract}




\maketitle

\section{INTRODUCTION}

Personalized recommender systems aim to predict items that align with individual users' preferences and underpin a wide range of online services~\cite{REC-SURVEY-1,REC-SURVEY-2,REC-SURVEY-3}. Collaborative filtering (CF)~\cite{CF-SURVEY,LightCCF} remains a foundational paradigm due to its effectiveness in learning user and item representations from interaction data. Recent advances in representation learning, including graph neural networks~\cite{PAAC} and contrastive learning~\cite{LightGCL,NLGCL}, have further improved CF by enhancing the quality of high-order structural representations. However, most existing methods implicitly assume homogeneous user preferences and overlook the fact that user decisions are driven by multiple interdependent latent intents across different granularity~\cite{SEM-VAE}. As a result, heterogeneous interaction signals are often entangled within learned representations, leading to limited robustness under sparse or noisy observations and reduced interpretability~\cite{DCCF,BIGCF}.

Decoupled representation learning addresses the limitations of monolithic embeddings and entangled interaction signals by factorizing heterogeneous interaction signals into independent latent dimensions, yielding interpretable and fine-grained intent-level representations. Recent studies explore this direction from complementary perspectives: \emph{DCCF}~\cite{DCCF} disentangles user intents to mitigate augmentation noise, \emph{BIGCF}~\cite{BIGCF} models bilateral intents at individual and collective levels, and \emph{IPCCF}~\cite{IPCCF} incorporates graph-aware intent propagation with contrastive alignment. Despite these advances, fundamental limitations remain in perspective-dependent interaction semantics and interaction-level intent alignment.

\textbf{\emph{Limitations of Single-Space Interaction Modeling.}} 
Most structural recommenders~\cite{AdaMCL,LightGCN,BIGCF,IPCCF,LightCCF,NR-GCF} adopt a representation-first paradigm, where user and item representations are learned from the interaction graph and user–item interactions are inferred via similarity-based evaluation. While effective at propagating structural information, such models explain interactions solely through global node representations, implicitly constraining all relations to a single interaction rule. In practice, interactions are inherently \emph{perspective-dependent}. From a \emph{user-centric} view~\cite{user-base-cf}, interactions reflect how latent user motivations activate specific item attributes; from an \emph{item-centric} view~\cite{item-base-cf}, they capture how intrinsic item semantics organize and differentiate user preferences. These perspectives induce \emph{asymmetric and non-interchangeable} interaction semantics. Consequently, explaining interactions with a single interaction rule typically entangles perspective-specific semantics, leading to semantic coupling and representational interference. Existing models therefore, despite sufficient representational capacity, remain limited in modeling fine-grained and perspective-consistent interactions driven by heterogeneous user motivations.

\textbf{\emph{Limitations in Interaction-Level Intent Alignment.}} 
State-of-the-art intent-aware models (e.g., \emph{BIGCF} and \emph{IPCCF}) typically integrate intent information with structural signals into unified node representations. Intent-related semantics are mainly regulated via interaction objectives defined over these mixed embeddings. Such objectives fail to explicitly constrain how corresponding user and item intent factors are activated or aligned within individual user--item interactions. Extra constraints like graph contrastive learning or intent-aware message propagation are often applied to intent embeddings. These act separately on user- and item-side embeddings, leaving explicit user–item intent alignment unaddressed. In practice, user intents are reflected through concrete interaction outcomes—i.e., consistent selection of intent-relevant items and rejection of irrelevant ones. Yet, current training objectives lack direct, interaction-level supervision that enforces the joint contribution of user and item intents on a per-interaction basis. As a result, even with well-structured node representations, the activation and coordination of intent dimensions within specific user--item pairs can remain underconstrained, especially under sparse settings.

These observations reveal a key limitation in collaborative filtering: effective intent modeling requires preserving perspective-dependent interaction semantics and promoting interaction-level intent alignment. To bridge this gap, we propose \emph{DMICF} (\textbf{D}ual-Perspective Disentangled \textbf{M}ulti-\textbf{I}ntent Alignment for Enhanced \textbf{C}ollaborative \textbf{F}iltering), a unified end-to-end framework that embodies these principles. \emph{DMICF} first constructs dual-perspective structural representations by jointly modeling high-order homophily and direct interactions from complementary user- and item-centric perspectives, preserving bilateral observational semantics. Built upon these representations, \emph{DMICF} employs a prototype-aware variational encoder that conditions inference on macro-level semantic prototypes, yielding uncertainty-aware and hierarchically structured latent intent representations. Finally, \emph{DMICF} explicitly aligns corresponding intent dimensions through interaction-driven supervision, ensuring that latent factors are directly grounded in observed user–item compatibility. Through this tightly integrated design, \emph{DMICF} coherently disentangles latent intents while enabling robust interaction prediction within a single optimization pipeline. Crucially, each component is flexible, with performance mainly determined by dual-perspective design and interaction-level alignment (Sec.~\ref{sec:abs_study}). We further provide a theoretical analysis suggesting that prototype-aware variational conditioning can help alleviate posterior collapse by inducing a structured aggregate posterior (Sec.~\ref{mitigating_intent_collapse}). Moreover, the reconstruction objective induces implicit intent-wise contrastive gradients that align positive user–item pairs along each latent intent dimension while repelling negatives, ensuring stable and semantically grounded intent alignment (Sec.~\ref{intent_alignment_analysis}). Our main contributions are as follows:
\begin{itemize}
    \item \textbf{Dual-Perspective Intent Modeling.} 
    We model user–item interactions from dual user- and item-centric perspectives, conditioning intent generation on perspective-specific structural signals to preserve perspective-dependent semantics.
    
    \item \textbf{Interaction-Level Intent Alignment.} 
    We propose \emph{DMICF}, an interaction-driven learning framework that enforces fine-grained, intent-wise alignment between users and items at the interaction level to preserve intent consistency.
    
    \item \textbf{Theoretical analysis and Empirical Validation.} 
    We offer theoretical insights into intent collapse mitigation and intent-wise optimization dynamics. Empirical results on multiple benchmarks show consistent improvements over strong baselines, with ablations validating each component.
\end{itemize}

\section{METHODOLOGY}

Let $\mathcal{U}= \left \{ u_{1},u_{2},\dots ,u_{I} \right \}$ and $\mathcal{V}= \left \{ v_{1},v_{2},\dots ,v_{J} \right \}$ denote the sets of users and items. User–item interactions are modeled as a bipartite graph $\mathcal{G}= \left \langle \left \{ \mathcal{U},\mathcal{I} \right \} ,\mathcal{E}\right \rangle$, where $\left ( u_{i},v_{j} \right )\in \mathcal{E}$ indicates an observed interaction. We encode $\mathcal{G}$ by a sparse binary matrix $\mathbf{A}\in \left \{ 0,1 \right \} ^{I\times J}$, where $\textbf{A}_{ij}=1$ iff $\left ( u_{i},v_{j} \right )\in \mathcal{E}$. The learning objective is to estimate the probability of edge existence $\prod_{u_{i}\in \mathcal{U}} \prod_{v_{j}\in \mathcal{V}}\mathbb{P} \left ( \textbf{Y}_{ij} \mid \mathbf{A} \right ) $.

\subsection{Model}

\begin{figure*}[t]
  \centering
  \includegraphics[width=\linewidth]{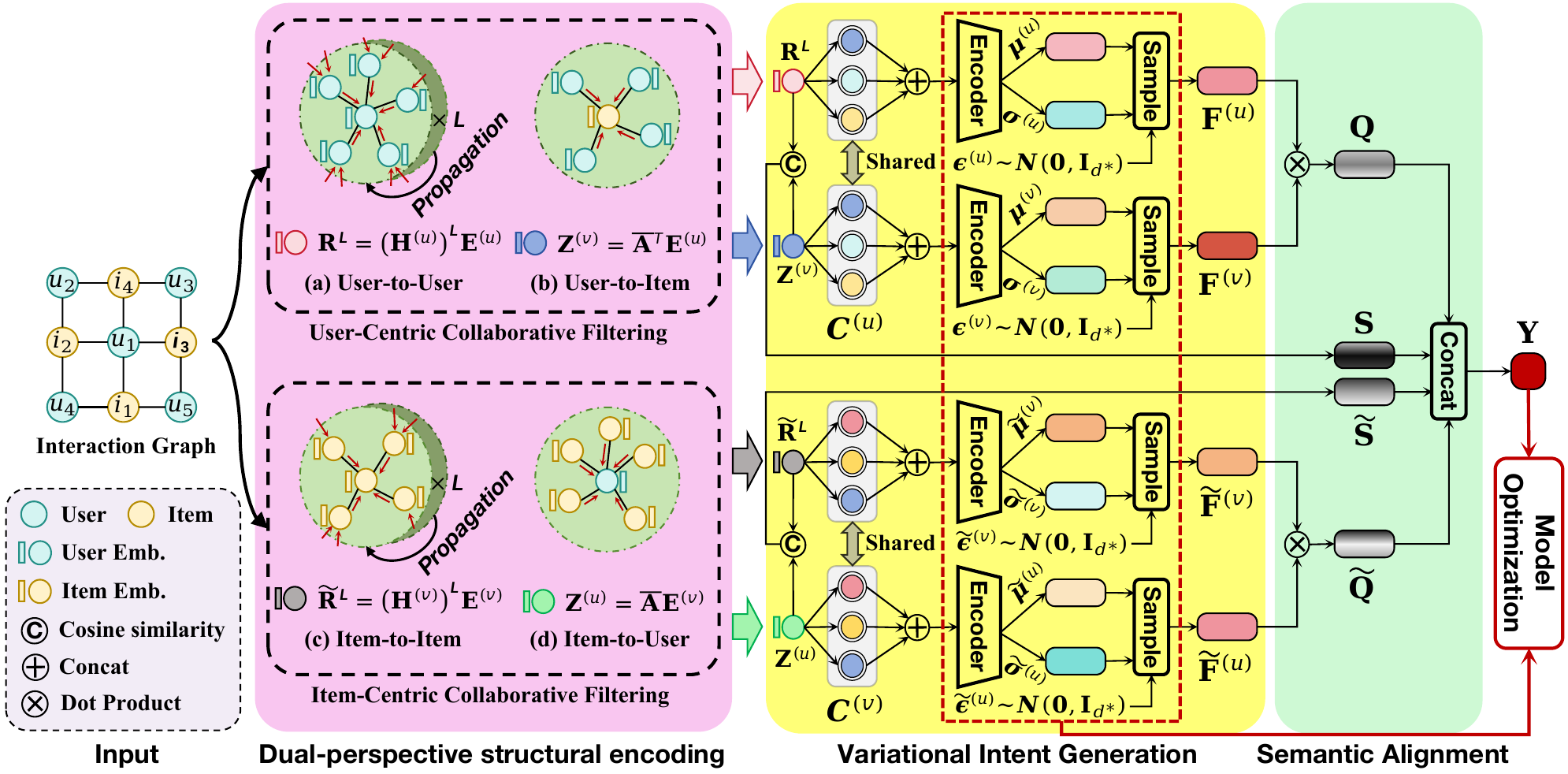}
  \caption{Overview of \emph{DMICF}, which preserves dual-perspective semantics and enables fine-grained intent alignment.}
  \label{fig:process}
\end{figure*}

As shown in Fig.~\ref{fig:process}, \emph{DMICF} consists of three tightly integrated modules: dual-perspective structural encoding, prototype-aware variational intent generation, and interaction-driven semantic alignment. Importantly, the performance gains of \emph{DMICF} primarily stem from its dual-perspective architecture and explicit intent alignment per perspective, rather than any particular module design choice (Sec.~\ref{sec:abs_study}). Accordingly, \textbf{\emph{each component is flexible and interchangeable}}, with core gains preserved across alternative designs.

\subsubsection{Dual-perspective structural encoding}
Most graph-based recommenders focus on user–item interactions, neglecting explicit user–user and item–item relations. While multi-hop propagation can capture high-order homogeneous signals, naive aggregation often introduces noise from weakly related nodes. We construct high-order user-user and item-item homogeneous graphs, denoted by $\textbf{H}^{\left ( u \right ) }$ and $\textbf{H}^{\left ( v \right ) }$, by computing Jaccard similarity over interaction neighborhoods with thresholding and ranking-based pruning~\cite{AdaMCL}:
\begin{align}
\label{eq:graph_uu}
\textbf{H}^{\left ( u \right ) }_{im} = \frac{\left | \mathcal{N}_{u_{i}} \cap \mathcal{N}_{u_{m}}  \right | }{\left | \mathcal{N}_{u_{i}} \cup \mathcal{N}_{u_{m}}  \right |}  &= \frac{\sum \textbf{A}_{i \cdot } \textbf{A}_{m \cdot }}{\sum \textbf{A}_{i \cdot } +  \sum \textbf{A}_{m \cdot} - \sum \textbf{A}_{i \cdot } \textbf{A}_{m \cdot }} \nonumber \\
\textbf{H}^{\left ( u \right ) }_{im} = 0 \; if \; \textbf{H}^{\left ( u \right ) }_{im} &< \eta \; or \; \textbf{H}^{\left ( u \right ) }_{im} \notin Top_{\omega} \left ( \textbf{H}^{\left ( u \right ) }_{i \cdot} \right ) \\
\textbf{H}^{\left ( v \right ) }_{jh} = \frac{\left | \mathcal{N}_{v_{j}} \cap \mathcal{N}_{v_{h}}  \right | }{\left | \mathcal{N}_{v_{j}} \cup \mathcal{N}_{v_{h}}  \right |}  &= \frac{\sum \textbf{A}_{\cdot j} \textbf{A}_{\cdot h}}{\sum \textbf{A}_{\cdot j} +  \sum \textbf{A}_{\cdot h } - \sum \textbf{A}_{\cdot j} \textbf{A}_{\cdot h}} \nonumber \\
\textbf{H}^{\left ( v \right ) }_{jh} = 0 \; if \; \textbf{H}^{\left ( v \right ) }_{jh} &< \tilde{\eta } \; or \; \textbf{H}^{\left ( v \right ) }_{jh} \notin Top_{\tilde{\omega }} \left ( \textbf{H}^{\left ( v \right ) }_{j \cdot} \right )
\end{align}
This pruning preserves strong homophily while suppressing noisy high-order connections. \emph{DMICF} encodes structural information from dual perspectives. From the user perspective, signals propagate among users via $L$-layer homogeneous message passing over $\textbf{H}^{\left ( u \right ) }$:
\begin{equation}
    \textbf{R}^{L} = \textbf{H}^{\left ( u \right ) } \textbf{R}^{L-1}=\dots =\left ( \textbf{H}^{\left ( u \right ) } \right ) ^{L} \textbf{E}^{\left ( u \right ) }
    \label{pro_uu}
\end{equation}
where $\textbf{E}^{\left ( u \right ) }\in \mathbb{R}^{I\times d} $ denotes initial user embeddings. For efficiency and stability, self-loops, feature transformations, and nonlinearities are omitted. Items aggregate user signals directly via observed interactions $\textbf{A}$, following a \emph{LightGCN}~\cite{LightGCN}-style formulation:
\begin{equation}
    \textbf{Z}^{\left ( v \right )} = \bar{\textbf{A} }^{T} \textbf{E}^{\left ( u \right ) } = \left ( \textbf{D}^{-\frac{1}{2} }_{\left ( u \right )} \textbf{A} \textbf{D}^{-\frac{1}{2} }_{\left ( v \right )} \right )^{T} \textbf{E}^{\left ( u \right ) }
\end{equation}
where $\textbf{D}_{\left ( u \right )} \in \mathbb{R}^{I\times I} $ and $\textbf{D}_{\left ( v \right )} \in \mathbb{R}^{J\times J} $ are diagonal degree matrices.

Symmetrically, from the item perspective, signals propagate among items via $\textbf{H}^{\left ( v \right ) } $ and users aggregate item signals through $\textbf{A}$:
\begin{align}
    \widetilde{\textbf{R}}^{L} =& \textbf{H}^{\left ( v \right ) } \widetilde{\textbf{R}}^{L-1} = \dots =\left ( \textbf{H}^{\left ( v \right ) } \right ) ^{L} \textbf{E}^{\left ( v \right ) }    \\
    \textbf{Z}^{\left ( u \right )} &= \bar{\textbf{A}} \textbf{E}^{\left ( v \right ) } = \left ( \textbf{D}^{-\frac{1}{2} }_{\left ( u \right )} \textbf{A} \textbf{D}^{-\frac{1}{2} }_{\left ( v \right )} \right ) \textbf{E}^{\left ( v \right ) }
\end{align}
where $\textbf{E}^{\left ( v \right ) }\in \mathbb{R}^{J\times d}$ denotes initial item embeddings. 

This dual-perspective design captures high-order homophily and direct interactions in a view-specific manner, producing structurally coherent user and item representations for intent modeling.

\subsubsection{Prototype-aware variational intent generation}
User–item interactions are driven by multiple latent intents. We introduce two sets of learnable semantic prototypes~\cite{BIGCF,IPCCF}, which serve as macro-level intent anchors from the user and item perspectives, respectively: $\mathcal{C}^{\left ( u \right ) } = \left \{ \textbf{c}^{\left ( u \right ) }_{k} \in \mathbb{R}^{d}  \right \} _{k=1}^{K}$ and $\mathcal{C}^{\left ( v \right ) } = \left \{ \textbf{c}^{\left ( v \right ) }_{k} \in \mathbb{R}^{d}  \right \} _{k=1}^{K}$. Structural embeddings are aligned with prototypes via cosine similarity to produce prototype-aware messages:
\begin{align}
    \textbf{T}^{\left ( u \right ) }_{i} =  \underset{k=1}{\overset{K}{\Vert}}
cos\left ( \textbf{R}^{L}_{i}, \textbf{c}^{\left ( u \right ) }_{k} \right ) \quad 
    \textbf{T}^{\left ( v \right ) }_{j} =  \underset{k=1}{\overset{K}{\Vert}}
cos\left ( \textbf{Z}^{\left ( v \right )}_{j}, \textbf{c}^{\left ( u \right ) }_{k} \right )   \nonumber \\
    \widetilde{\textbf{T}}^{\left ( u \right ) }_{i} =  \underset{k=1}{\overset{K}{\Vert}}
cos\left ( \textbf{Z}^{\left ( u \right )}_{i}, \textbf{c}^{\left ( v \right ) }_{k} \right ) \quad
    \widetilde{\textbf{T}}^{\left ( v \right ) }_{j} =  \underset{k=1}{\overset{K}{\Vert}}
cos\left ( \widetilde{\textbf{R}}^{L}_{j}, \textbf{c}^{\left ( v \right ) }_{k} \right )
    \label{eq:global_stru}
\end{align}
Concatenating similarities over all $K$ prototypes yields macro-level intent representations, with each dimension corresponds to an independent latent factor. Conditioned on dual-perspective structural encodings, behaviorally similar nodes exhibit consistent prototype responses, enabling disentangled and interpretable intent modeling.

To capture uncertainty and diversity in latent intents, we employ a variational encoder that maps prototype-aware messages into a multivariate Gaussian space. Lightweight MLPs $Encoder_{\mu}$ and $Encoder_{log \, \sigma^{2} }$ parameterize the mean and log-variance in $\mathbb{R}^{d^{\ast}}$:
\begin{align}
    \textbf{M}^{\left ( u \right ) }_{i} = Encoder_{\mu}\left ( \textbf{T}^{\left ( u \right ) }_{i} \right )
    \quad
    \textbf{V}^{\left ( u \right ) }_{i} = Encoder_{log \, \sigma^{2} }\left ( \textbf{T}^{\left ( u \right ) }_{i} \right ) \nonumber \\
    \textbf{M}^{\left ( v \right ) }_{j} = Encoder_{\mu}\left ( \textbf{T}^{\left ( v \right ) }_{j} \right )
    \quad
    \textbf{V}^{\left ( v \right ) }_{j} = Encoder_{log \, \sigma^{2} }\left ( \textbf{T}^{\left ( v \right ) }_{j} \right )
\end{align}

\begin{align}
    \widetilde{\textbf{M}}^{\left ( u \right ) }_{i} = Encoder_{\mu}\left ( \widetilde{\textbf{T}}^{\left ( u \right ) }_{i} \right )
    \quad
    \widetilde{\textbf{V}}^{\left ( u \right ) }_{i} = Encoder_{log \, \sigma^{2} }\left ( \widetilde{\textbf{T}}^{\left ( u \right ) }_{i} \right ) \nonumber \\
    \widetilde{\textbf{M}}^{\left ( v \right ) }_{j} = Encoder_{\mu}\left ( \widetilde{\textbf{T}}^{\left ( v \right ) }_{j} \right )
    \quad
    \widetilde{\textbf{V}}^{\left ( v \right ) }_{j} = Encoder_{log \, \sigma^{2} }\left ( \widetilde{\textbf{T}}^{\left ( v \right ) }_{j} \right )
\end{align}
Latent intent embeddings are sampled via the reparameterization trick with $\epsilon^{(u)}, \epsilon^{(v)}, \widetilde{\epsilon}^{(u)}, \widetilde{\epsilon}^{(v)} \sim \mathcal{N}(\textbf{0}, \textbf{I}_{d^{\ast}})$, $\textbf{I}_{d^{\ast}}$ the identity matrix:
\begin{align}
    \textbf{F}^{\left ( u \right ) }_{i} = \textbf{M}^{\left ( u \right ) }_{i} + \mathrm{e}^{\frac{1}{2}  \textbf{V}^{\left ( u \right ) }_{i}}  \odot \epsilon^{\left ( u \right ) } \quad
    \textbf{F}^{\left ( v \right ) }_{j} = \textbf{M}^{\left ( v \right ) }_{j} + \mathrm{e}^{\frac{1}{2} \textbf{V}^{\left ( v \right ) }_{j}} \odot \epsilon^{\left ( v \right ) } \nonumber \\
    \widetilde{\textbf{F}}^{\left ( u \right ) }_{i} = \widetilde{\textbf{M}}^{\left ( u \right ) }_{i} + \mathrm{e}^{ \frac{1}{2} \widetilde{\textbf{V}}^{\left ( u \right ) }_{i}} \odot \widetilde{\epsilon}^{\left ( u \right ) } \quad
    \widetilde{\textbf{F}}^{\left ( v \right ) }_{j} = \widetilde{\textbf{M}}^{\left ( v \right ) }_{j} + \mathrm{e}^{ \frac{1}{2} \widetilde{\textbf{V}}^{\left ( v \right ) }_{j}} \odot \widetilde{\epsilon}^{\left ( v \right ) }
\end{align}

To regularize variational intent distributions and promote structured disentanglement, we impose a \textbf{\emph{KL divergence regularization}} that aligns each approximate posterior with a standard Gaussian prior $p\left ( \textbf{g} \right)=\mathcal{N}\left ( \textbf{0}, \textbf{I}_{d^{\ast}} \right )$. For an observed interaction $(u_{i}, v_{j})$,
\begin{align}
    \label{eq:KL loss}
    \mathcal{L}_{KL}^{\left ( i,j \right ) } 
    &= \mathcal{D}_{KL}\left ( \textbf{b}_{i}^{\left ( u \right ) } \parallel p\left ( \textbf{g} \right )  \right )  
    + \mathcal{D}_{KL}\left ( \textbf{b}_{j}^{\left ( v \right ) } \parallel p\left ( \textbf{g} \right )  \right )   \nonumber \\
    &+ \mathcal{D}_{KL}\left ( \widetilde{\textbf{b}}_{i}^{\left ( u \right ) } \parallel p\left ( \textbf{g} \right )  \right )  
    + \mathcal{D}_{KL}\left ( \widetilde{\textbf{b}}_{j}^{\left ( v \right ) } \parallel p\left ( \textbf{g} \right )  \right ) 
\end{align}
where the approximate posterior distributions are given by
\begin{align}
    \textbf{b}_{i}^{(u)} = \mathcal{N}\left ( \textbf{M}_{i}^{\left ( u \right ) } ,diag(\mathrm {e} ^{\textbf{V}_{i}^{\left ( u \right ) }}  )  \right )  
    \;
    \textbf{b}_{j}^{\left ( v \right ) } = \mathcal{N}\left ( \textbf{M}_{j}^{\left ( v \right ) } ,diag(\mathrm {e} ^{\textbf{V}_{j}^{\left ( v \right ) }} )  \right ) \nonumber \\
    \widetilde{\textbf{b}}_{i}^{\left ( u \right ) } = \mathcal{N}\left ( \widetilde{\textbf{M}}^{\left ( u \right ) }_{i} ,diag(\mathrm {e} ^{ \widetilde{\textbf{V}}^{\left ( u \right ) }_{i}})  \right )  
    \;
    \widetilde{\textbf{b}}_{j}^{\left ( v \right ) } = \mathcal{N}\left ( \widetilde{\textbf{M}}^{\left ( v \right ) }_{j} ,diag(\mathrm {e} ^{\widetilde{\textbf{V}}^{\left ( v \right ) }_{j}})  \right )
\end{align}
The KL divergence between a diagonal Gaussian posterior and the standard Gaussian prior has the closed-form:
\begin{equation}
    \mathcal{D}_{KL}\left ( \mathcal{N} \left ( \textbf{m},diag\left ( \mathrm {e}^{\textbf{v}}\right ) \right ) \parallel p\left ( \textbf{g} \right)   \right ) = - \frac{1}{2}\sum_{l=1}^{d^{\ast}}\left ( 1+ \textbf{v}_{l}- \textbf{m}_{l}^{2} - \mathrm {e}^{\textbf{v}_{l} }     \right )
\end{equation}
Regularizing across dual perspectives encourages factorized latent semantics while preserving uncertainty. Negative samples are excluded from KL computation to avoid constraining unobserved or noisy interactions, ensuring regularization is applied only to informative behavioral signals.

\subsubsection{Interaction-driven semantic alignment}
User–item interactions are driven by intent-level compatibility between user preferences and item semantics. To capture such fine-grained correspondence, we perform \textbf{dimension-wise semantic alignment} between prototype-aware latent intent embeddings. Specifically, intent compatibility is measured via element-wise interactions between corresponding user and item intent dimensions:
\begin{align}
    \textbf{Q}_{ij} = Alignment\left ( \textbf{F}^{\left ( u \right ) }_{i}, \textbf{F}^{\left ( v \right ) }_{j}\right ) = \textbf{F}^{\left ( u \right ) }_{i} \odot \textbf{F}^{\left ( v \right ) }_{j}  \\
    \widetilde{\textbf{Q}}_{ij} = Alignment\left ( \widetilde{\textbf{F}}^{\left ( u \right ) }_{i}, \widetilde{\textbf{F}}^{\left ( v \right ) }_{j} \right ) = \widetilde{\textbf{F}}^{\left ( u \right ) }_{i} \odot \widetilde{\textbf{F}}^{\left ( v \right ) }_{j}
    \label{eq:intent_alignment}
\end{align}
These intent-wise interactions preserve factor-level semantics and enable explicit alignment at the latent dimension level. The effect is evaluated via Sec.~\ref{intent_alignment_analysis} and the $w/o\,HP$ variant. Lightweight MLPs are applied to model nonlinear dependencies among intent factors:
\begin{equation}
    \textbf{Q}_{ij} = \phi_{MLP} \left ( \textbf{Q}_{ij} \right ) \in \mathbb{R} ^{d^{\star}}
    \quad
    \widetilde{\textbf{Q}}_{ij} = \phi_{MLP} \left ( \widetilde{\textbf{Q}}_{ij} \right ) \in \mathbb{R} ^{d^{\star}}
\end{equation}
In parallel, we compute auxiliary structural similarity signals by measuring user–item similarity within user- and item-centric structural embedding spaces, respectively:
\begin{equation}
    \textbf{S}_{ij}= cos\left ( \textbf{R}^{L}_{i}, \textbf{Z}^{\left ( v \right )}_{j} \right ) 
    \quad
    \widetilde{\textbf{S}}_{ij}= cos\left ( \textbf{Z}^{\left ( u \right )}_{i}, \widetilde{\textbf{R}}^{L}_{j} \right )  
\end{equation}
Finally, intent alignment and structural similarity from dual perspectives are fused to produce the interaction representation:
\begin{equation}
    \textbf{Y}_{ij} = AGG\left ( \textbf{Q}_{ij}, \widetilde{\textbf{Q}}_{ij},\textbf{S}_{ij},\widetilde{\textbf{S}}_{ij}\right ) = \phi_{MLP}\left ( \textbf{Q}_{ij} \parallel \widetilde{\textbf{Q}}_{ij} \parallel \textbf{S}_{ij} \parallel \widetilde{\textbf{S}}_{ij}\right ) 
\end{equation}
which jointly encodes fine-grained intent compatibility and structural consistency for interaction prediction.

\subsubsection{Model Optimization}

\emph{DMICF} is trained by jointly optimizing reconstruction accuracy and latent regularization. Given an observed interaction $\left ( u_{i}, v_{j} \right )$, we adopt negative sampling by randomly drawing $n$ unobserved items $\mathcal{N}^{neg}_{u_{i}}$ from the item set $\mathcal{V}$. Interaction scores are normalized via a temperature-scaled softmax:
\begin{equation}
    \label{eq:softmax}
    \textbf{P}_{ij} = \frac{\mathrm{exp} \left ( \textbf{Y}_{ij}/\tau  \right ) }{\mathrm{exp} \left ( \textbf{Y}_{ij}/\tau  \right )+{\textstyle \sum_{v_{l}\in \mathcal{N}^{neg}_{u_{i}} }} \mathrm{exp} \left ( \textbf{Y}_{il}/\tau  \right ) } 
\end{equation}
where $\tau$ denotes the temperature parameter controlling distribution sharpness. The reconstruction loss is defined as
\begin{equation}
    \label{eq:rec loss}
    \mathcal{L}_{rec} =  {\textstyle \sum_{\left ( i,j \right ) \in \mathcal{E}^{train}}} \left [ \left ( 1 - \textbf{P}_{ij}\right )^{2}+  {\textstyle \sum_{v_{l}\in \mathcal{N}^{neg}_{u_{i}}}} \left ( 0- \textbf{P}_{il} \right)^{2}\right ] 
\end{equation}
The overall objective combines reconstruction and variational regularization, with $\lambda$ balances predictive performance and disentangled intent modeling:
\begin{equation}
    \mathcal{L}= \mathcal{L}_{rec}+  \lambda \ast \mathbb{E}_{\left ( i,j \right ) \in \mathcal{E}^{train} } \left [ \mathcal{L}_{KL}^{\left ( i,j \right ) }  \right ] 
\end{equation}

\begin{remark} [\textbf{Optimization Insight.}]
    By coupling dimension-wise intent alignment with variational regularization, the reconstruction objective induces fine-grained, interaction-driven specialization of each latent intent dimension, while the KL term preserves structured diversity across the latent space. This joint optimization ensures that intent representations emerge collaboratively from user–item interactions, rather than being independently shaped by global dispersion or instance-level contrastive constraints.
\end{remark}

\subsection{Model Analysis}
\subsubsection{Time Complexity Analysis} 

We fix the high-order propagation depth to $L=1$. Each variational encoder consists of two two-layer MLP branches, $Encoder_{\mu}$ and $Encoder_{log \, \sigma^{2} }$, each of size $\left [ K,d_{1},d^{\ast} \right ]$. The intent alignment module adopts two three-layer MLPs of size $\left [d^{\ast},d_{2},d_{3},d^{\star} \right ]$, while the dual-perspective fusion is implemented by a lightweight MLP of size $\left [2\times d^{\star} + 2,d_{4},1 \right]$.

Computing enriched node embeddings costs $\mathcal{O}\left ( 2\left | \mathcal{E}  \right |d+  \left ( I+  J \right )Sd \right ) $. Prototype relevance scoring and variational intent encoding introduce $\mathcal{O}\left ( \left ( I+ J \right ) Kd \right )  $ and $\mathcal{O}\left ( \left ( I+J \right ) \left ( Kd_{1}+  d_{1}d^{\ast } \right )  \right ) $, respectively. Dual-perspective alignment scale as $\mathcal{O}\left ( 2\left | \mathcal{E} \right | \left ( d^{\ast }d_{2}+ d_{2}d_{3}+ d_{3}d^{\star}\right ) \right )  $, and the fusion adds $\mathcal{O}\left ( 2\left | \mathcal{E} \right |d^{\star}d_{4} \right )$. Overall, the dominant complexity of \emph{DMICF} grows linearly with the number of observed interactions $\left | \mathcal{E}  \right |$, ensuring scalability on large graphs.

\subsubsection{Mitigating Intent Collapse.}
\label{mitigating_intent_collapse}
We offer a theoretical analysis to help explain why prototype-aware conditioning may alleviate posterior collapse. For clarity, we analyze user intents under the user perspective; the same arguments extend symmetrically to all remaining cases. Empirical evidence is reported in Sec.~\ref{intent_anti_collapse}.

\begin{proposition}\label{proposition}[\textbf{Aggregate Posterior as Implicit Mixture}] Let $\mathbf{o}^{\left ( u \right ) }_{i}$ denote the latent intent variable of user $u_{i}$. The variational posterior $q_{\phi}(\mathbf{o}^{\left ( u \right ) }_{i} | \textbf{T}^{\left ( u \right ) }_{i})$ captures individual-level intent uncertainty, while the aggregate posterior $q_{agg}(\mathbf{o}^{\left ( u \right ) }) = \mathbb{E}_{p_{data}(u_{i})}[q_{\phi}(\mathbf{o}^{\left ( u \right ) }|u_{i})]$ characterizes population-level intent semantics. In standard VAEs, $q_{agg}(\mathbf{o}^{\left ( u \right ) })$ often collapses to the unimodal prior $\mathcal{N}\left ( \textbf{0},\textbf{I} \right )$. In contrast, by conditioning inference on the prototype-aware macro-level representation $\textbf{T}^{\left ( u \right ) }_{i}$ (Eq.\ref{eq:global_stru}), DMICF induces an aggregate posterior that approximates a mixture distribution anchored by semantic prototypes:
\begin{align}
    q_{agg}(\mathbf{o}^{(u)}) \approx \sum_{k=1}^{K} \int & p_{data}(u_{i}) \cdot \mathbb{P}(\mathcal{C}^{(u)}_k \mid u_i) \cdot \nonumber \\
    &\mathcal{N}\left(\mathbf{o}^{(u)}; \boldsymbol{\mu}_\phi(\mathbf{T}^{(u)}_i), diag(\boldsymbol{\sigma}^2_\phi(\mathbf{T}^{(u)}_i))\right)du_i
\end{align}
where $\mathbb{P}(\mathcal{C}^{(u)}_k \mid u_{i})$ denotes the soft assignment of user $u_{i}$ to prototype $k$, implicitly encoded within $\textbf{T}^{(u)}_{i}$. Consequently, the latent space is geometrically structured as a union of localized semantic regions rather than a single convex region, preserving intent diversity.
\end{proposition}

We consider the information-theoretic decomposition of the evidence lower bound (ELBO)~\cite{ELBO} to substantiate the anti-collapse property. Let $\mathbf{a}_{i} \in \left \{ 0,1 \right \}^{J}$ denote the interaction vector of $u_{i}$, $p (\mathbf{o}^{\left ( u \right ) }) =\mathcal{N}\left ( \textbf{0}, \textbf{I}_{d^{\ast}} \right )$ be the standard Gaussian prior. The expectation of the objective over the empirical data distribution $p_{data}(u_{i})$ is:
\begin{align}
    \mathbb{E}_{u_i}\left [ \mathcal{L}_{ELBO} \right ]  = & \mathbb{E}_{u_i, q}\left [ \log p\left ( \mathbf{a}_{i} \mid \mathbf{o}^{\left ( u \right ) }_{i} \right ) \right ]  - \beta \cdot I\left ( \mathbf{o}^{\left ( u \right ) }_{i}; u_{i} \right ) \nonumber \\
    & - \beta \cdot \mathcal{D}_{KL}\left ( q_{agg}(\mathbf{o}^{\left ( u \right ) })   \parallel p(\mathbf{o}^{\left ( u \right ) } ) \right )
\end{align}
where $\mathbb{E}_{u_i, q}$ denotes the joint expectation over $p_{data}\left ( u_{i} \right )$ and the variational posterior $q_{\phi }(\mathbf{o}^{\left ( u \right ) }_{i} \mid \mathbf{a}_{i})$. $I(\mathbf{o}^{\left ( u \right ) }_{i}; u_{i} )$ is the mutual information between the user index and the latent code. Our Macro–Micro architecture mitigates collapse through two coordinated effects:

\textbf{High-Entropy Conditioning.} Posterior collapse typically occurs when $I(\mathbf{o}^{(u)}_{i}; u_{i})\to 0$. Conditioning inference on the prototype-aware representation $\textbf{T}^{(u)}_{i}$, instead of sparse identifiers, encourages $q_{\phi}(\mathbf{o}^{(u) }_{i} \mid \textbf{T}^{(u)}_{i})$ to encode high-entropy semantic signals. As a result, the latent code remains dependent on the input semantics.

\textbf{Structured Manifold Regularization.} The marginal KL term $\mathcal{D}_{KL}(q_{agg}(\mathbf{o}^{(u) }) \parallel p(\mathbf{o}^{(u)}))$ aligns the population posterior with the unimodal prior, while prototype alignment (Eq.\ref{eq:global_stru}) biases individual means $\boldsymbol{\mu}_\phi (\textbf{T}^{(u)}_{i})$ toward distinct semantic regions in the latent space. This equilibrium effectively renders the KL term a \textit{local compactness constraint}, encouraging a structured latent decomposition in which macro-prototypes specify region locations, while micro-level variational noise captures within-region uncertainty.

\subsubsection{Optimization Analysis of Intent Alignment} 
\label{intent_alignment_analysis}
We analyze how the reconstruction loss induces contrastive, intent-wise updates: positive items are aligned with users along each latent intent dimension, while negative items are pushed in the opposite direction.

For clarity, we assume a batch size of 1 and consider a single observed interaction $\left ( u_{i}, v_{j}   \right ) $ with sampled negatives. We examine the gradient with respect to the $\kappa$-th dimension of the latent item intent representation $\textbf{F}^{\left ( v \right ) }_{j}\left [ \kappa  \right ]$. By the chain rule,
\begin{align}
    \frac{\partial \mathcal{L}_{rec}}{\partial \textbf{F}^{\left ( v \right ) }_{j}\left [ \kappa  \right ]} & = \frac{\partial \mathcal{L}_{rec}}{\partial \textbf{P}_{ij}} \cdot \frac{\partial \textbf{P}_{ij}}{\partial \textbf{Y}_{ij}} \cdot \frac{\partial \textbf{Y}_{ij}}{\partial \textbf{F}^{\left ( v \right ) }_{j}\left [ \kappa  \right ]} \nonumber \\
    & \propto - \left ( 1- \textbf{P}_{ij} \right ) \cdot \textbf{P}_{ij}\left ( 1- \textbf{P}_{ij} \right ) \cdot \textbf{F}^{\left ( u \right ) }_{i}\left [ \kappa  \right ]
\end{align}
Since all scalar factors are positive except the leading minus sign, gradient descent updates $\textbf{F}^{\left ( v \right ) }_{j}\left [ \kappa  \right ]$ in the direction of $\textbf{F}^{\left ( u \right ) }_{i}\left [ \kappa  \right ]$. Consequently, positive interactions explicitly align the $\kappa$-th item intent with the corresponding user intent. For a negative item $v_{l}\in \mathcal{N}^{neg}_{u_{i}}$:
\begin{align}
    \frac{\partial \mathcal{L}_{rec}}{\partial \textbf{F}^{\left ( v \right ) }_{l}\left [ \kappa  \right ]} \propto + \textbf{P}_{il}\cdot \left ( 1- \textbf{P}_{il} \right )\cdot \textbf{F}^{\left ( u \right ) }_{i}\left [ \kappa  \right ] 
\end{align}
which drives $\textbf{F}^{\left ( u \right ) }_{i}\left [ \kappa  \right ]$ away from the corresponding intent direction.

As $\textbf{Y}_{ij}$ integrates intent-level compatibility and structural similarity in an end-to-end differentiable manner, intent-wise gradients are jointly backpropagated across users and items, yielding consistent and semantically grounded intent alignment. Sec.~\ref{case_study} provides a qualitative case study.

\subsubsection{Comparative Implications.} 

Recent intent-aware recommenders, particularly contrastive learning–based methods, mitigate posterior collapse through instance-level objectives that promote uniform utilization of latent intents (e.g., \emph{BIGCF}). While effective, such regularization implicitly flattens distinct intent dimensions into a shared semantic manifold, reducing discriminability. In contrast, \emph{DMICF} employs prototype-aware variational conditioning to organize the aggregate posterior into localized semantic regions, inducing a coarse-to-fine latent intent space~\cite{varium,SEM-VAE}. Prototype conditioning provides shared semantic anchors, while interaction-level refinement captures personalized deviations. Moreover, reconstruction-driven gradients in \emph{DMICF} induce implicit, dimension-wise intent alignment, enabling fine-grained semantic specialization. However, existing CF objectives typically rely on instance-level margin or similarity supervision, operating on global scalar similarity and thus failing to induce consistent dimension-level semantics.

\section{EXPERIMENTS}
We conduct extensive experiments to evaluate \emph{DMICF}. We benchmark it against state-of-the-art recommenders in terms of accuracy, efficiency, and robustness under varying interaction regimes. We further perform ablation and qualitative analyses to validate key design components and empirically substantiate our theoretical claims on intent alignment and collapse mitigation.

\subsection{Experimental Setting}

\subsubsection{\textbf{Datasets}}

\begin{table}
  \caption{Statistics of the Experimental Datasets.}
  \label{tab:dataset}
  \begin{tabular}{ccccc}
    \toprule
    Dataset & \# user  & \# item  & \# interactions  & sparsity \\
    \midrule
    \textbf{Amazon}  & 78,578 & 77,801 & 2,240,156 & 99.96\%  \\
    \textbf{Tmall}  & 47,939 & 41,390 & 2,357,450 & 99.88\%  \\
    \textbf{ML-10M}  & 69,878 & 10,195 & 6,999,171 & 99.02\%  \\
  \bottomrule
\end{tabular}
\end{table}

\begin{table*}[t]
  \caption{Overall performance comparisons on Amazon, Tmall and ML-10M datasets w.r.t. Recall@N (abbreviated as R@N) and
NDCG@N (abbreviated as N@N). \textbf{OOM} denotes out-of-memory failure during training.}
  \label{tab:commands}
  \renewcommand{\arraystretch}{1.2}
  \resizebox{\textwidth}{!}{%
  \begin{tabular}{c|cccccc|cccccc|cccccc}
    \toprule
    \multirow{2}{*}{Model} 
      & \multicolumn{6}{c|}{Amazon} 
      & \multicolumn{6}{c|}{Tmall} 
      & \multicolumn{6}{c}{ML-10M} \\
      \cline{2-19}
      & R@10 & N@10 & R@20 & N@20 & R@40 & N@40 
      & R@10 & N@10 & R@20 & N@20 & R@40 & N@40
      & R@10 & N@10 & R@20 & N@20 & R@40 & N@40 \\
    \hline
    IMP-GCN
    & 0.0513 & 0.0531 & 0.0787 & 0.0617 & 0.1172 & 0.0744 
    & 0.0410 & 0.0367 & 0.0658 & 0.0467 & 0.1017 & 0.0592 
    & 0.1451 & 0.2590 & 0.2343 & 0.2711 & 0.3429 & 0.2985 \\
    LightGCN
    & 0.0525 & 0.0542 & 0.0827 & 0.0638 & 0.1246 & 0.0776 
    & 0.0396 & 0.0354 & 0.0628 & 0.0449 & 0.0982 & 0.0571
    & 0.1563 & 0.2735 & 0.2482 & 0.2862 & 0.3586 & 0.3145 \\
    DirectAU 
    & 0.0559 & 0.0576 & 0.0877 & 0.0677 & 0.1321 & 0.0823 
    & 0.0444 & 0.0397 & 0.0707 & 0.0503 & 0.1074 & 0.0630 
    & 0.1577 & 0.2735 & 0.2506 & 0.2872 & 0.3614 & 0.3160 \\
    HCCF 
    & 0.0531 & 0.0547 & 0.0813 & 0.0636 & 0.1209 & 0.0766 
    & 0.0436 & 0.0389 & 0.0695 & 0.0495 & 0.1078 & 0.0628 
    & 0.1516 & 0.2679 & 0.2427 & 0.2801 & 0.3510 & 0.3072 \\
    LightGCL 
    & 0.0624 & 0.0639 & 0.0975 & 0.0751 & 0.1468 & 0.0913 
    & 0.0462 & 0.0411 & 0.0733 & 0.0521 & 0.1133 & 0.0660 
    & 0.1602 & 0.2823 & 0.2382 & 0.2837 & 0.3264 & 0.3016 \\
    AdaMCL
    & 0.0586 & 0.0598 & 0.0898 & 0.0691 & 0.1342 & 0.0838 
    & 0.0451 & 0.0400 & 0.0716 & 0.0509 & 0.1104 & 0.0643 
    & 0.1672 & 0.2791 & 0.2616 & 0.2904 & 0.3735 & 0.3190 \\
    EGCF 
    & 0.0606 & 0.0628 & 0.0924 & 0.0729 & 0.1357 & 0.0870 
    & 0.0465 & 0.0417 & 0.0738 & 0.0528 & 0.1112 & 0.0658 
    & 0.1748 & 0.3013 & 0.2717 & 0.3084 & 0.3865 & 0.3359 \\
    LightCCF 
    & 0.0707 & 0.0727 
    & \textcolor{blue}{\underline{0.1087}} & 0.0847 
    & \textcolor{blue}{\underline{0.1593}} & 0.1012
    & \textcolor{blue}{\underline{0.0515}} 
    & \textcolor{blue}{\underline{0.0456}} 
    & \textcolor{blue}{\underline{0.0821}}
    & \textcolor{blue}{\underline{0.0581}}
    & \textcolor{blue}{\underline{0.1258}}
    & \textcolor{blue}{\underline{0.0733}}
    & 0.1800 
    & \textcolor{blue}{\underline{0.3097}} 
    & 0.2790 & 0.3193 & 0.3960 & 0.3472 \\
    MacridVAE
    & 0.0545 & 0.0559 & 0.0833 & 0.0649 & 0.1239 & 0.0783 
    & 0.0426 & 0.0378 & 0.0685 & 0.0483 & 0.1042 & 0.0607 
    & 0.1643 & 0.2903 & 0.2613 & 0.3030 & 0.3774 & 0.3319 \\
    DCCF 
    & 0.0561 & 0.0577 & 0.0863 & 0.0661 & 0.1296 & 0.0804 
    & 0.0440 & 0.0391 & 0.0702 & 0.0486 & 0.1098 & 0.0642 
    & \textbf{OOM} & \textbf{OOM}
    & \textbf{OOM} & \textbf{OOM} 
    & \textbf{OOM} & \textbf{OOM}  \\
    BIGCF 
    & 0.0641 & 0.0648 & 0.0986 & 0.0759 & 0.1460 & 0.0915 
    & 0.0465 & 0.0409 & 0.0742 & 0.0523 & 0.1148 & 0.0663 
    & 0.1776 & 0.3095 & 0.2764 & 0.3188 & 0.3928 & 0.3457 \\
    IPCCF 
    & \textcolor{blue}{\underline{0.0739}}
    & \textcolor{blue}{\underline{0.0771}}
    & 0.1083
    & \textcolor{blue}{\underline{0.0875}}
    & 0.1530
    & \textcolor{blue}{\underline{0.1020}}
    & 0.0503 & 0.0450 & 0.0789 & 0.0566 & 0.1186 & 0.0703 
    & \textcolor{blue}{\underline{0.1814}}
    & 0.3094 
    & \textcolor{blue}{\underline{0.2841}}
    & \textcolor{blue}{\underline{0.3243}}
    & \textcolor{blue}{\underline{0.4058}}
    & \textcolor{blue}{\underline{0.3561}}  \\
    \textbf{DMICF}
    & \cellcolor[HTML]{FADBD8} \textbf{0.0777}
    & \cellcolor[HTML]{FADBD8} \textbf{0.0792}
    & \cellcolor[HTML]{FADBD8} \textbf{0.1175}
    & \cellcolor[HTML]{FADBD8} \textbf{0.0918}
    & \cellcolor[HTML]{FADBD8} \textbf{0.1716}
    & \cellcolor[HTML]{FADBD8} \textbf{0.1096}
    & \cellcolor[HTML]{FADBD8} \textbf{0.0588}
    & \cellcolor[HTML]{FADBD8} \textbf{0.0518}
    & \cellcolor[HTML]{FADBD8} \textbf{0.0930}
    & \cellcolor[HTML]{FADBD8} \textbf{0.0658}
    & \cellcolor[HTML]{FADBD8} \textbf{0.1411}
    & \cellcolor[HTML]{FADBD8} \textbf{0.0825}
    & \cellcolor[HTML]{FADBD8} \textbf{0.1934}
    & \cellcolor[HTML]{FADBD8} \textbf{0.3418}
    & \cellcolor[HTML]{FADBD8} \textbf{0.3048}
    & \cellcolor[HTML]{FADBD8} \textbf{0.3548}
    & \cellcolor[HTML]{FADBD8} \textbf{0.4331}
    & \cellcolor[HTML]{FADBD8} \textbf{0.3850}  \\
    \emph{impr.\%}
    & \textcolor{red}{\textbf{5.14}} $\textcolor{red}{\uparrow}$
    & \textcolor{red}{\textbf{2.72}} $\textcolor{red}{\uparrow}$
    & \textcolor{red}{\textbf{8.10}} $\textcolor{red}{\uparrow}$
    & \textcolor{red}{\textbf{4.91}} $\textcolor{red}{\uparrow}$
    & \textcolor{red}{\textbf{7.72}} $\textcolor{red}{\uparrow}$
    & \textcolor{red}{\textbf{7.45}} $\textcolor{red}{\uparrow}$
    
    & \textcolor{red}{\textbf{14.17}} $\textcolor{red}{\uparrow}$
    & \textcolor{red}{\textbf{13.60}} $\textcolor{red}{\uparrow}$
    & \textcolor{red}{\textbf{13.28}} $\textcolor{red}{\uparrow}$
    & \textcolor{red}{\textbf{13.25}} $\textcolor{red}{\uparrow}$
    & \textcolor{red}{\textbf{12.16}} $\textcolor{red}{\uparrow}$
    & \textcolor{red}{\textbf{12.55}} $\textcolor{red}{\uparrow}$
    
    & \textcolor{red}{\textbf{6.62}} $\textcolor{red}{\uparrow}$
    & \textcolor{red}{\textbf{10.36}} $\textcolor{red}{\uparrow}$
    & \textcolor{red}{\textbf{7.29}} $\textcolor{red}{\uparrow}$
    & \textcolor{red}{\textbf{9.40}} $\textcolor{red}{\uparrow}$
    & \textcolor{red}{\textbf{6.73}} $\textcolor{red}{\uparrow}$
    & \textcolor{red}{\textbf{8.12}} $\textcolor{red}{\uparrow}$  \\
    
    \bottomrule
  \end{tabular}%
  }
\end{table*}

We evaluate \emph{DMICF} on three widely used benchmarks—Amazon, Tmall, and ML-10M—covering diverse scales, domains, and sparsity levels. Table~\ref{tab:dataset} summarizes their key statistics. For fair comparison, we adopt the preprocessing pipeline and data splits released by \emph{LightGCL}~\cite{LightGCL}. Model performance is assessed with standard top-$N$ metrics: Recall@N, which measures the coverage of relevant items, and NDCG@N, which evaluates ranking quality.

\subsubsection{\textbf{Baselines}}

We compare \emph{DMICF} against representative state-of-the-art recommenders, grouped by paradigm: (i) \textbf{GNN-based}: \emph{IMP-GCN}~\cite{IMP-GCN}, \emph{LightGCN}, \emph{DirectAU}~\cite{DirectAU}; 
(ii) \textbf{SSL-based}: \emph{HCCF}~\cite{HCCF}, \emph{LightGCL}~\cite{LightGCL}, \emph{AdaMCL}~\cite{AdaMCL}, \emph{EGCF}~\cite{EGCF}, \emph{LightCCF}~\cite{LightCCF}; 
(iii) \textbf{Intent modeling-based}: \emph{MacridVAE}~\cite{MacridVAE}, \emph{DCCF}, \emph{BIGCF}, \emph{IPCCF}. All baselines use publicly available implementations, with hyperparameters set to reported best configurations or tuned via grid search.

\subsubsection{\textbf{Hyperparameter Setting}}

\emph{DMICF} is implemented in PyTorch. Following prior work~\cite{HCCF,LightGCL}, we use an \textbf{embedding dimension of 32, batch size of 4096}, and the Adam optimizer with Xavier initialization. For all datasets, the number of intent prototypes is set to $K=48$. The variational encoders ($Encoder_{\mu}$ and $Encoder_{log \, \sigma^{2} }$), intent alignment MLPs, and dual-perspective fusion MLP are instantiated as $[48,32,80]$, $[80,128,64,16]$, and $[34,32,1]$, respectively. The softmax temperature is $\tau=0.2$ and the KL weight is $\lambda=100$. For homogeneous graph construction, we set $\eta = \tilde{\eta}=0.2$ and $\omega=\tilde{\omega}=10$ on Amazon and Tmall, while applying stricter filtering ($\eta=0.8$, $\tilde{\eta}=0.2$, $\omega=20$, and $\tilde{\omega}=10$) on the denser ML-10M dataset. Early stopping is applied with a patience of 3 epochs.

\subsection{Performance Comparisons}

\subsubsection{\textbf{Overall Comparisons}}
Table~\ref{tab:commands} compares \emph{DMICF} with representative baselines. We make the following observations.

(1) \emph{DMICF} consistently achieves the best performance across datasets and metrics. In particular, it improves Recall@20 over the strongest baseline by 8.10\%, 13.28\%, and 7.29\% on Amazon, Tmall, and ML-10M, respectively, demonstrating robust gains across varying data scales and sparsity levels.
(2) \emph{DMICF} consistently outperforms existing intent-disentanglement baselines (e.g., \emph{DCCF}, \emph{BIGCF}, and \emph{IPCCF}), highlighting the effectiveness of grounding fine-grained intents through explicit interaction-level alignment.
(3) \emph{DMICF} substantially outperforms both SSL-based and GNN-based models. Compared with SSL-based approaches that separate positive and negative pairs via instance-level contrastive learning, \emph{DMICF} induces contrastive updates through reconstruction, aligning or repelling interactions along each latent intent dimension. This dimension-wise contrastive effect enforces consistent semantic specialization and enables finer-grained separation. Compared with GNN-based models relying on local aggregation, \emph{DMICF} captures more comprehensive structural signals via dual-perspective modeling and conditions intent learning on both direct interactions and higher-order homophily, yielding stronger structural expressiveness, especially under sparse and long-tail regimes.

\subsubsection{\textbf{Training Time and Convergence}}
\begin{figure}[t]
  \centering
  \includegraphics[width=\linewidth]{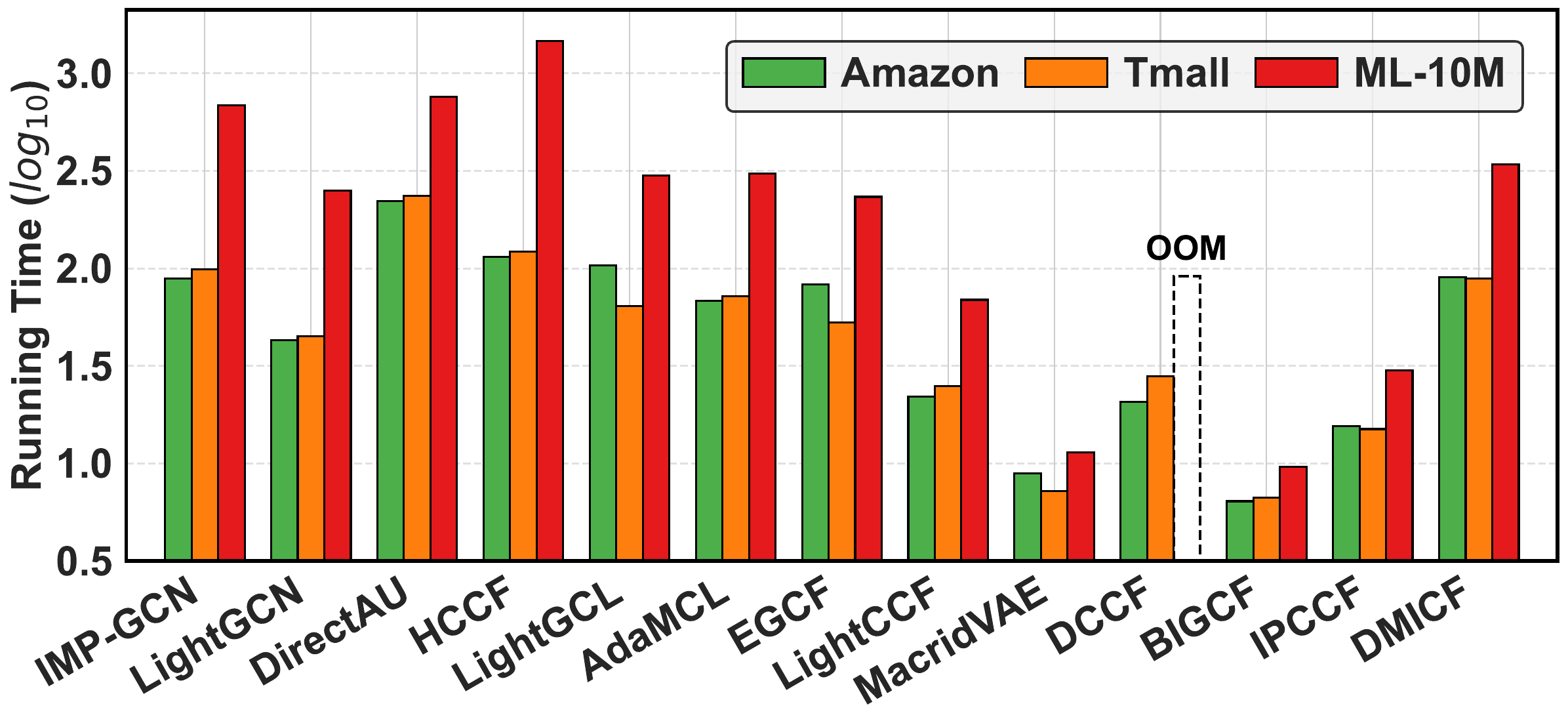}
  \caption{Per-epoch training time of \emph{DMICF} versus baselines.}
  \label{fig:runtime}
\end{figure}

\begin{figure}[t]
  \centering
  \includegraphics[width=\linewidth]{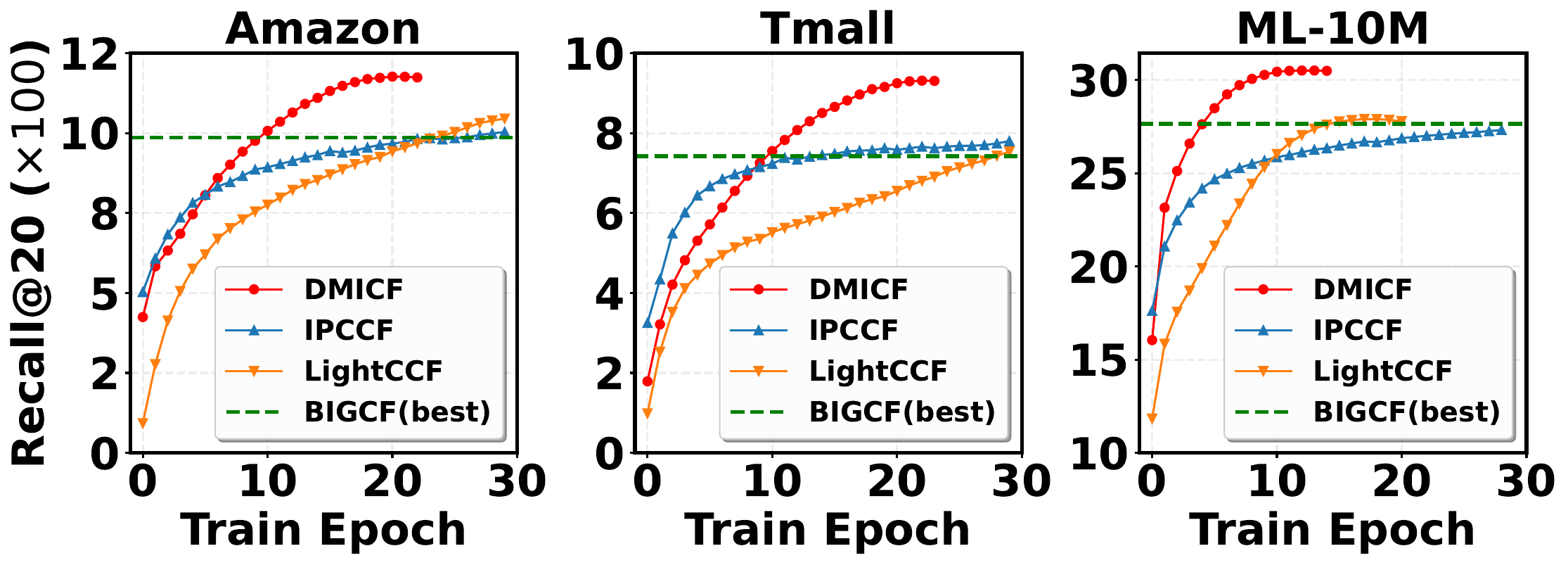}
  \caption{Early-stage performance (first 30 epochs) of \emph{DMICF}, \emph{IPCCF}, and \emph{LightCCF}. Slower models are omitted for clarity.}
  \label{fig:run_process}
\end{figure}

Fig.~\ref{fig:runtime} compares per-epoch training time across models under a unified setting, excluding data loading and system overhead. Empirically, \emph{DMICF} exhibits near-linear scaling with respect to the number of interactions. Its per-epoch cost is higher than that of \emph{MacridVAE}, \emph{DCCF}, \emph{BIGCF}, and \emph{IPCCF}, which rely on fixed-size user sampling, but remains comparable to other mainstream models, notably \emph{LightGCL}. Despite a slightly higher per-epoch overhead, Fig.~\ref{fig:run_process} shows that \emph{DMICF} converges substantially faster, reaching the peak performance of strong baselines such as \emph{LightCCF} and \emph{IPCCF} in fewer epochs. This efficiency gain likely stems from its ability to capture higher-order relational signals and enforce early-stage intent alignment, thereby operating with a broader effective receptive field.

\subsubsection{\textbf{Impact of Interaction Frequency}} 
\begin{figure}[t]
  \centering
  \includegraphics[width=\linewidth]{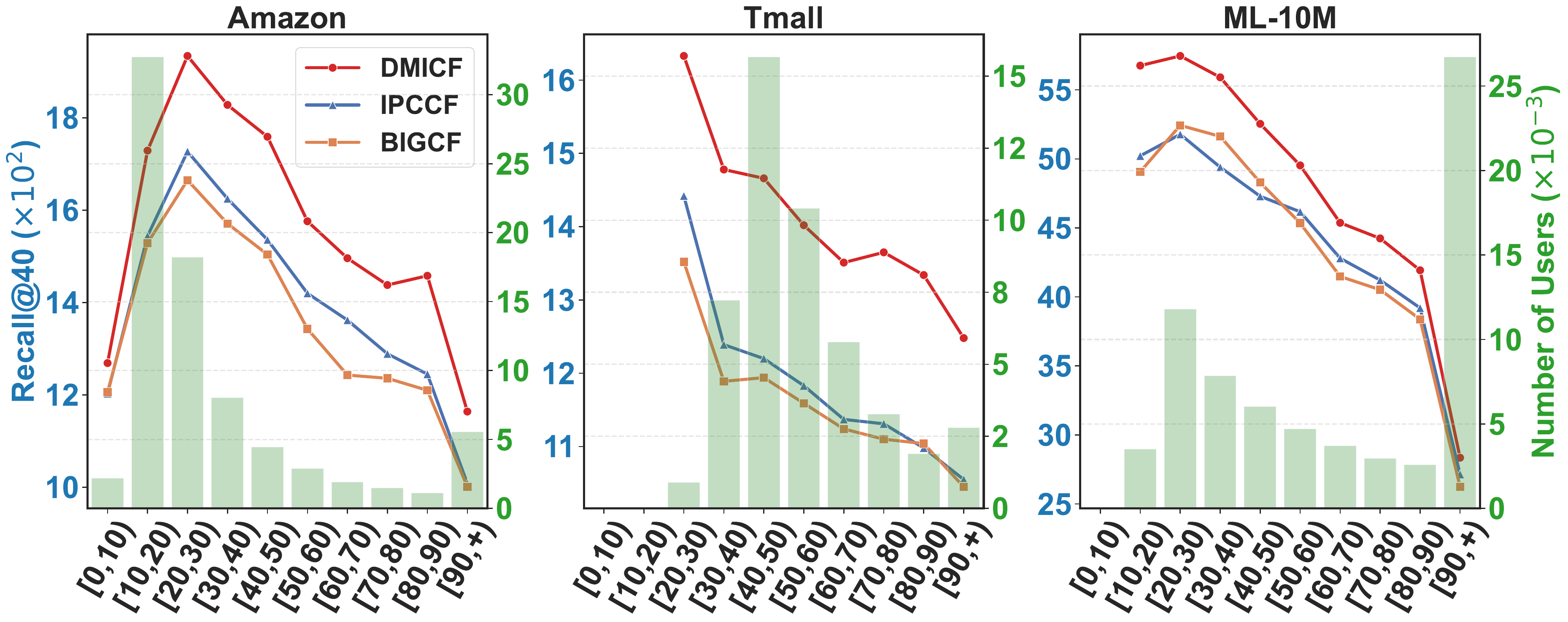}
  \caption{Performance of \emph{DMICF}, \emph{IPCCF} and \emph{BIGCF} across user groups binned by interaction frequency. Tmall spans $\left [ 0,20 \right ) $, and ML-10M has no users in $\left [ 0,10 \right ) $.}
  \label{fig:user_group}
\end{figure}

To analyze model behavior across interaction frequency regimes, we partition test users into ten groups based on interaction frequency and evaluate performance within each group. For clarity, we report \emph{DMICF} and the strongest intent-aware baselines, \emph{IPCCF} and \emph{BIGCF}. Fig.~\ref{fig:user_group} illustrates distinct interaction distributions, with peaks around $\left [ 10,20 \right ) $ on Amazon, $\left [ 40,50 \right ) $ on Tmall, and $\left [ 90,+ \right ) $ on ML-10M. Across all datasets and all frequency subsets, \emph{DMICF} consistently achieves the best performance, with particularly pronounced gains on Tmall. On datasets where extremely sparse users exist (e.g., the $[0,10)$ group in Amazon), all methods exhibit performance degradation; however, \emph{DMICF} maintains clear advantages over \emph{IPCCF} and \emph{BIGCF}, indicating stronger sparsity resistance. We further observe that moderately dense interactions (e.g., $[20,30)$) often yield peak performance, as structural evidence becomes sufficient while user intents remain relatively coherent. In contrast, overly dense regimes can impair performance as user intents become increasingly heterogeneous. Nevertheless, \emph{DMICF} remains consistently superior, suggesting a stronger ability to capture intent diversity with higher discriminative power. Overall, these results demonstrate that modeling user–item interactions from dual perspectives enables richer exploitation of both direct interactions and high-order structural dependencies, while prototype-aware variational intent alignment at the interaction level yields more robust and discriminative preference representations across diverse interaction regimes.

\subsection{In-depth Studies of DMICF}

\subsubsection{\textbf{Ablation Studies}}
\label{sec:abs_study}
We examine the contribution of each component in \emph{DMICF} by comparing it with eight variants:

(1) $w/o \,SS$, remove the auxiliary structural-similarity signals $\textbf{S}_{ij}$ and $\widetilde{\textbf{S}}_{ij}$;
(2) $w/o \,HOS$, remove high-order homogeneous propagation by setting $L=0$;
(3) $w/o \,VIC$, removes variational sampling by replacing stochastic intent representations with their mean vectors, e.g., $\textbf{F}^{\left ( u \right ) }_{i} = \textbf{M}^{\left ( u \right ) }_{i}$;
(4) $w/o \,KL$, remove KL regularization by setting $\lambda=0$;
(5) $w/o \,IP$, remove prototype guidance by replacing prototype responses with structural embeddings, e.g., $\textbf{T}^{\left ( u \right ) }=\textbf{R}^{L}$;
(6) $w/o \,HP$, replaces dimension-wise intent alignment via Hadamard product with feature concatenation, e.g., $\textbf{Q}_{ij} = \phi_{MLP}(\textbf{F}^{\left ( u \right ) }_{i} \parallel  \textbf{F}^{\left ( v \right ) }_{j}) $; 
(7) $w/o \,IA$, removes explicit interaction-level intent alignment while retaining intent representations. Structural embeddings and intent features are directly fused at the node level within each perspective prior to prediction (e.g., $\mathbf{R}^{L} + \mathbf{F}^{(u)}$ for users and $\mathbf{Z}^{(v)} + \mathbf{F}^{(v)}$ for items in the user perspective), followed by dual-perspective edge semantic fusion. 
(8) $w/o \,Dual$, adopts a single GNN pipeline over the user–item bipartite graph with matched embedding dimensionality and comparable model capacity to \emph{DMICF}. Structural embeddings obtained via graph message passing are used to generate variational intent representations, followed by dimension-wise intent alignment. Interaction modeling relies on a unified edge semantic, without perspective-specific interaction semantics. 

Table~\ref{tab:abs study} reports the results, revealing the following observations. 
\textbf{(i) Robustness to individual component choices.} 
Variants that preserve both dual-perspective interaction modeling and interaction-level intent alignment—while modifying individual components such as structural similarity ($w/o\,SS$), high-order structure ($w/o \,HOS$), variational intent encoding ($w/o\,VIC$), KL regularization ($w/o\,KL$), prototype guidance ($w/o\,IP$), or replacing Hadamard Product with concatenation ($w/o \,HP$)—exhibit only mild and consistent performance drops. Except for $w/o \,IP$, all variants consistently outperform strong structural baselines (e.g., \emph{BIGCF}), indicating that \emph{DMICF}'s improvements do not rely on specific module instantiations as long as the core interaction semantics are preserved. The larger degradation under $w/o\,IP$ arises from intent collapse (Sec.~\ref{intent_anti_collapse}), where latent intents lose semantic separation without prototype anchoring. The performance drop of $w/o \,HOS$ decreases as interaction density increases, suggesting that strong direct bipartite interactions allow interaction-level intent alignment to compensate for missing high-order same-type signals. $w/o \,HP$ exhibits a modest drop, as the Hadamard product encodes dimension-wise correspondences between user and item intents, although alternative interaction functions (concatenation, dot product, or attention) remain compatible without substantially affecting performance. 
\textbf{(ii) High sensitivity to interaction semantics.} 
The most substantial drops are observed for $w/o\,Dual$ and $w/o\,IA$. Removing interaction-level intent alignment consistently degrades performance, despite preserving intent-aware feature enrichment, indicating that alignment is essential for inducing consistent intent semantics across users and items. Performance also degrades sharply under $w/o \,Dual$ (e.g., a $\sim$36\% Recall@20 drop on Amazon), despite matched embedding dimensionality and comparable model capacity, indicating that the degradation stems from the loss of perspective-dependent interaction semantics rather than reduced capacity. Overall, the results show that \emph{DMICF}'s gains primarily arise from well-defined and aligned interaction semantics, while most architectural components mainly affect robustness and stability.

\begin{table}[t]
\centering
\caption{Ablation results of \emph{DMICF} on benchmark datasets.}
\resizebox{\columnwidth}{!}{%
\begin{tabular}{c|cc|cc|cc}
\hline
\multirow{2}{*}{Model} & \multicolumn{2}{c|}{Amazon} & \multicolumn{2}{c|}{Tmall} & \multicolumn{2}{c}{ML1-0M} \\
& R@20 & N@20 & R@20 & N@20 & R@20 & N@20 \\
\hline
w/o SS  
& \textcolor{blue}{\underline{0.1142}} 
& \textcolor{blue}{\underline{0.0894}} 
& \textcolor{blue}{\underline{0.0917}} 
& \textcolor{blue}{\underline{0.0650}} 
& \textcolor{blue}{\underline{0.3028}} 
& \textcolor{blue}{\underline{0.3514}} \\
w/o HOS & 0.1062 & 0.0809 & 0.0881 & 0.0615 & 0.3017 & 0.3506 \\
w/o VIC & 0.1089 & 0.0835 & 0.0894 & 0.0622 & 0.3012 & 0.3499 \\
w/o KL  & 0.1022 & 0.0782 & 0.0833 & 0.0592 & 0.2825 & 0.3327 \\
w/o IP  & 0.0721 & 0.0555 & 0.0808 & 0.0575 & 0.2884 & 0.3356 \\
w/o HP  & 0.1116 & 0.0878 & 0.0885 & 0.0613 & 0.2998 & 0.3472 \\
w/o IA  & 0.0938  & 0.0726 & 0.0797  & 0.0570  & 0.2804  & 0.3222 \\
w/o Dual  & 0.0864  & 0.0705  & 0.0781 & 0.0561 & 0.2743 & 0.3161 \\
\emph{DMICF} 
& \cellcolor[HTML]{FADBD8} 0.1175
& \cellcolor[HTML]{FADBD8} 0.0918 
& \cellcolor[HTML]{FADBD8} 0.0930
& \cellcolor[HTML]{FADBD8} 0.0658 
& \cellcolor[HTML]{FADBD8} 0.3048
& \cellcolor[HTML]{FADBD8} 0.3548 \\
\hline
\end{tabular}%
}
\label{tab:abs study}
\end{table}

\begin{figure}[t]
  \centering
  \includegraphics[width=\linewidth]{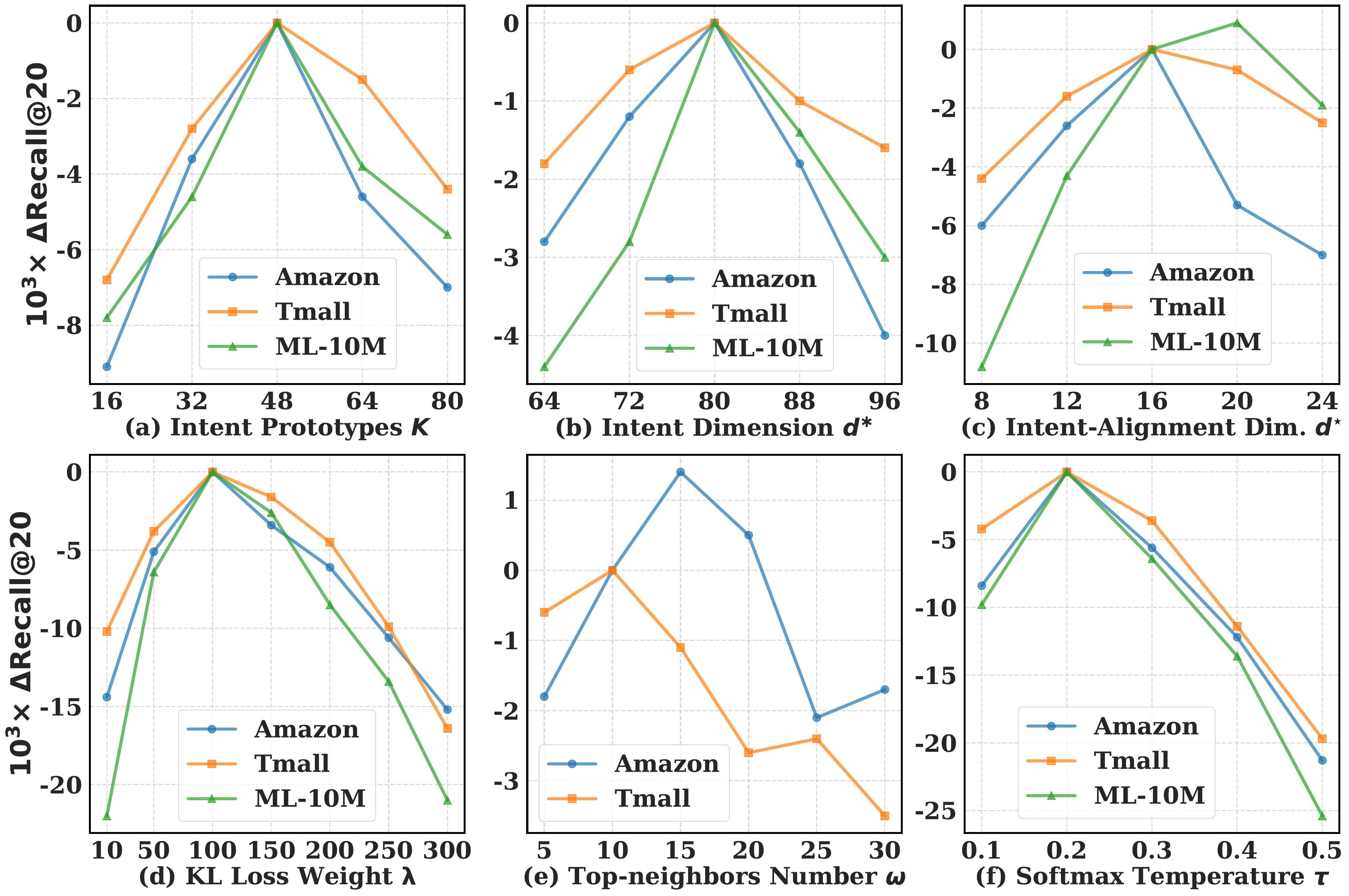}
  \caption{Hyperparameter study of \emph{DMICF}.}
  \label{fig:params}
\end{figure}

\subsubsection{\textbf{Hyperparameter Sensitivities}}
Fig.~\ref{fig:params} studies the sensitivity of \emph{DMICF} to key hyperparameters. 
Fig.~\ref{fig:params}(a) shows a rise-then-fall trend as the number of intent prototypes increases, indicating that while more prototypes improve expressiveness, excessive ones introduce redundancy and noise under interaction-level supervision. 
Fig.~\ref{fig:params}(b) demonstrates moderate sensitivity to the latent dimension $d^{\ast}$: small values limit expressiveness, whereas overly large ones reduce representational distinctiveness. Fig.~\ref{fig:params}(c) reveals that the preferred intent-alignment dimension $d^\star$ depends on data density: sparse datasets favor smaller values, while denser datasets benefit from higher capacity to capture richer intent interactions. 
Fig.~\ref{fig:params}(d) highlights a trade-off in the KL weight $\lambda$: insufficient regularization leads to unstable intents, while overly large values restrict personalization. 
Fig.~\ref{fig:params}(e) shows that \emph{DMICF} is relatively insensitive to $\omega$($=\tilde{\omega}$): larger values benefit sparser graphs by providing stronger structural priors, but may introduce noise in denser settings. 
Finally, Fig.~\ref{fig:params}(f) indicates that the temperature $\tau$ has a notable impact, with $\tau=0.2$ consistently yielding strong performance across datasets. 
Overall, a largely shared hyperparameter configuration yields strong performance across all datasets, requiring only minor tuning of $\omega$ and $d^\star$, to which \emph{DMICF} is relatively insensitive.

\subsubsection{\textbf{Effectiveness of Higher-Order Message Propagation}}

We analyze the effect of higher-order structural propagation by varying the message passing depth $L$ in \emph{DMICF}. Table~\ref{tab:layers} reports performance under different depths. \emph{DMICF} achieves the best results at $L=1$, while deeper propagation enlarges the receptive field but introduces over-smoothing that degrades node discriminability. Setting $L=0$ ($w/o \,HOS$) yields a clear performance drop, indicating that higher-order propagation provides essential structural context for intent modeling. Notably, even without higher-order propagation ($L=0$), \emph{DMICF} remains competitive with strong intent-aware baselines, highlighting that its intent modeling and matching mechanisms are inherently robust. On ML-10M, the performance gain from $L=0$ to $L=1$ is relatively small, suggesting that dense interaction graphs already provide sufficient local structural signals, reducing the marginal benefit of deeper propagation.

\begin{figure}[t]
  \centering
  \includegraphics[width=\linewidth]{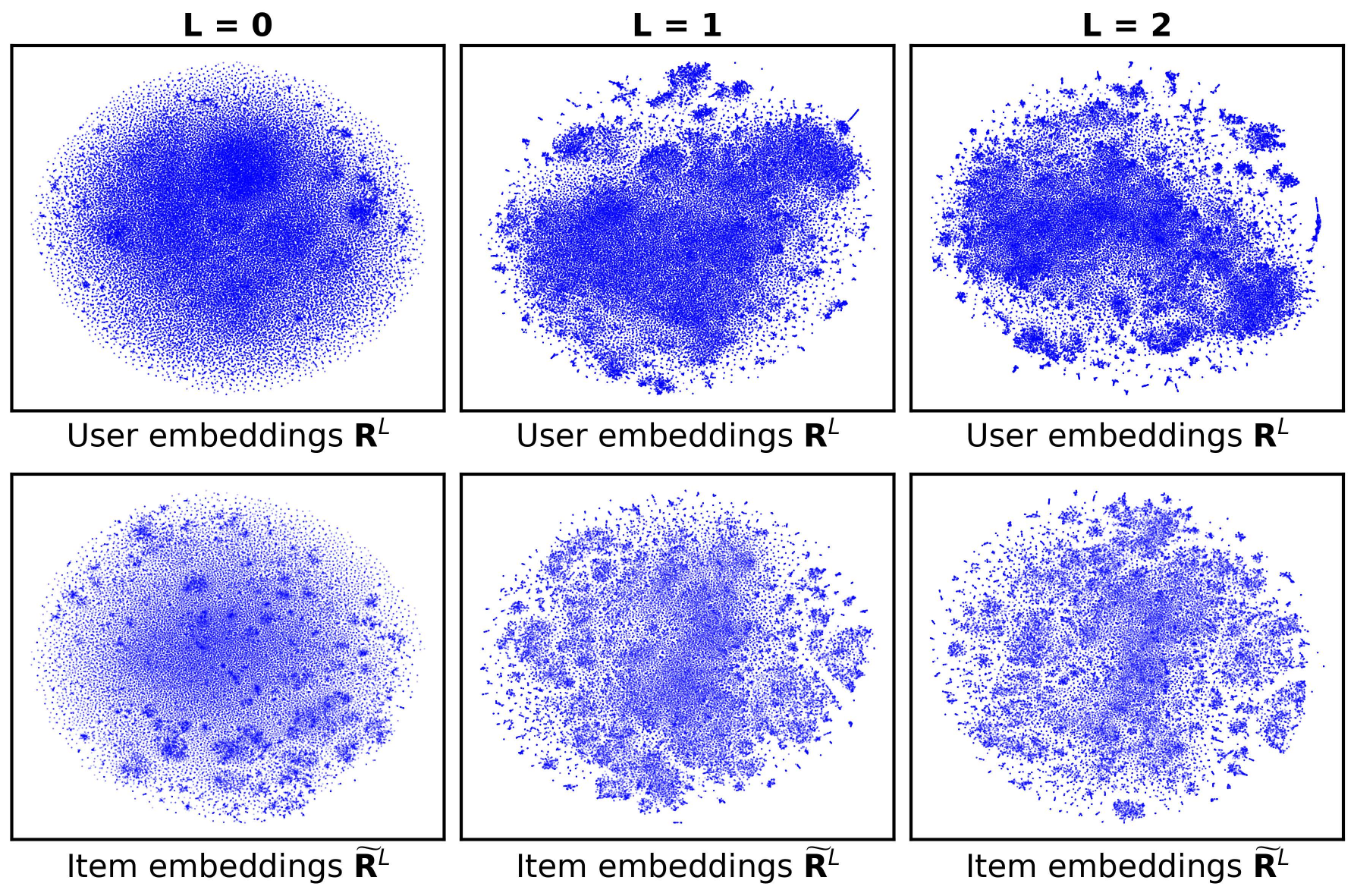}
  \caption{t-SNE visualization of higher-order aggregated embeddings $\textbf{R}^{L}$ and $\widetilde{\textbf{R}}^{L}$ under varying propagation depths $L$.}
  \label{fig:user_item_embedding}
\end{figure}

Fig.~\ref{fig:user_item_embedding} visualizes user and item embeddings on Amazon under different propagation depths using t-SNE. With $L=0$, embeddings exhibit weak structural organization, while $L=1$ produces well-separated clusters, reflecting effective aggregation of homophilous interactions. Increasing to $L=2$ yields slightly blurred boundaries, consistent with over-smoothing. These visualizations validate the quantitative results, indicating that moderate higher-order propagation is critical for learning compact yet expressive representations.

\subsubsection{\textbf{Case Study}}
\label{case_study}
\begin{figure}[t]
  \centering
  \includegraphics[width=\linewidth]{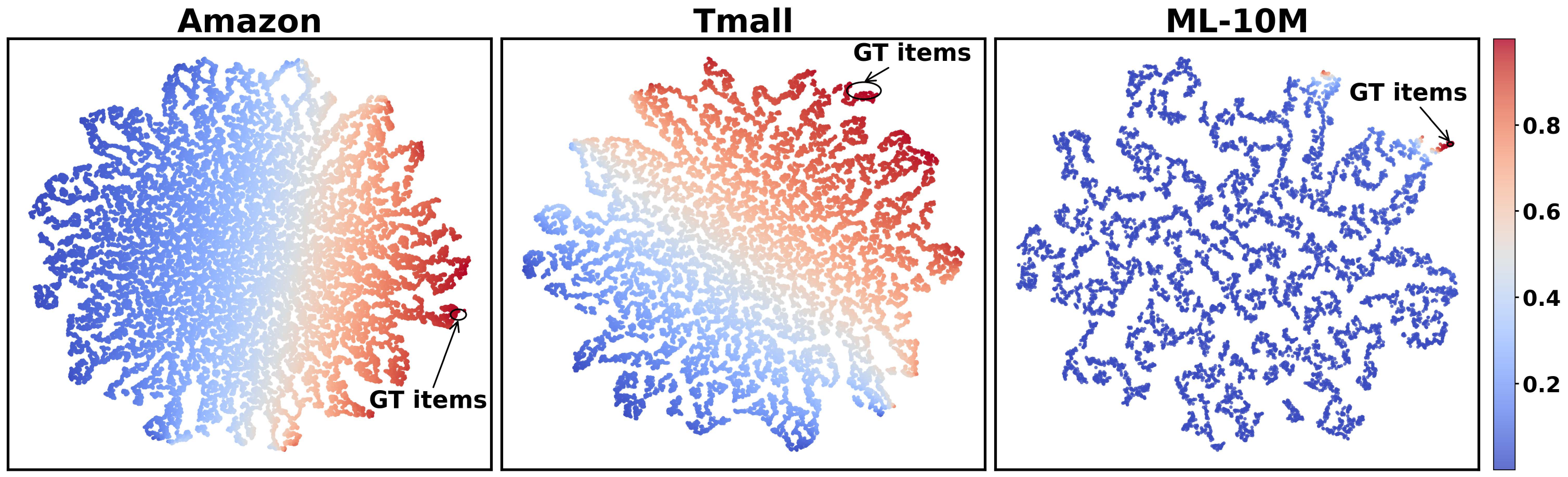}
  \caption{t-SNE visualization of interaction predictions for selected users on Amazon ($u_{3501}$, HR@20 = 0.8696), Tmall ($u_{15435}$, HR@20 = 0.7083), and ML-10M ($u_{54253}$, HR@20 = 0.9048).}
  \label{fig:link_feature}
\end{figure}

Fig.~\ref{fig:link_feature} visualizes predicted interaction scores for a representative user over all items across datasets in the intent-aligned interaction space $\mathbf{X}_{ij} = \mathbf{Q}_{ij} \parallel \widetilde{\mathbf{Q}}_{ij}$. Each point corresponds to a user-item pair and is colored by its predicted score $\mathbf{Y}_{ij}$, where the spatial distribution reflects interaction-level intent compatibility. On Amazon and Tmall, scores vary smoothly across the space, forming coherent high-score regions. Ground-truth (GT) items consistently fall within or near these regions, while non-interacted items are distributed in lower-score areas. This indicates that the learned representation organizes items according to their compatibility with the target user in a structured and semantically consistent manner. In contrast, on ML-10M, high-score items concentrate in highly localized neighborhoods, with most of the space assigned low scores. GT items cluster tightly within these localized regions, suggesting sharper and more selective user–item compatibility patterns on this dataset. Across all datasets, GT items exhibit clear separation trends from negative items in both spatial location and predicted score. These observations are consistent with the intent-wise contrastive optimization in Sec.~\ref{intent_alignment_analysis}, which pulls positive interactions into compatible regions while pushing negatives away.

\begin{table}[t]
\centering
\caption{\emph{DMICF} performance w.r.t. propagation depth $L$.}
\begin{tabular}{c|cc|cc|cc}
\hline
\multirow{2}{*}{} & \multicolumn{2}{c|}{Amazon} & \multicolumn{2}{c|}{Tmall} & \multicolumn{2}{c}{ML-10M} \\
& R@20 & N@20 & R@20 & N@20 & R@20 & N@20 \\
\hline
$L=0$ & 0.1062 & 0.0809 & 0.0881 & 0.0615 & 0.3017 & 0.3506 \\
$L=1$ 
& \cellcolor[HTML]{FADBD8}  0.1175
& \cellcolor[HTML]{FADBD8}  0.0918
& \cellcolor[HTML]{FADBD8}  0.0930
& \cellcolor[HTML]{FADBD8}  0.0658
& \cellcolor[HTML]{FADBD8}  0.3048
& \cellcolor[HTML]{FADBD8}  0.3548 \\
$L=2$  
& \textcolor{blue}{\underline{0.1154}} 
& \textcolor{blue}{\underline{0.0901}}
& \textcolor{blue}{\underline{0.0913}}
& \textcolor{blue}{\underline{0.0647}}
& \textcolor{blue}{\underline{0.3036}}
& \textcolor{blue}{\underline{0.3522}} \\
$L=3$  & 0.1142 & 0.0901 & 0.0902 & 0.0640 & 0.3030 & 0.3512 \\
\hline
\end{tabular}%
\label{tab:layers}
\end{table}

\begin{figure*}[t]
  \centering
  \includegraphics[width=\linewidth]{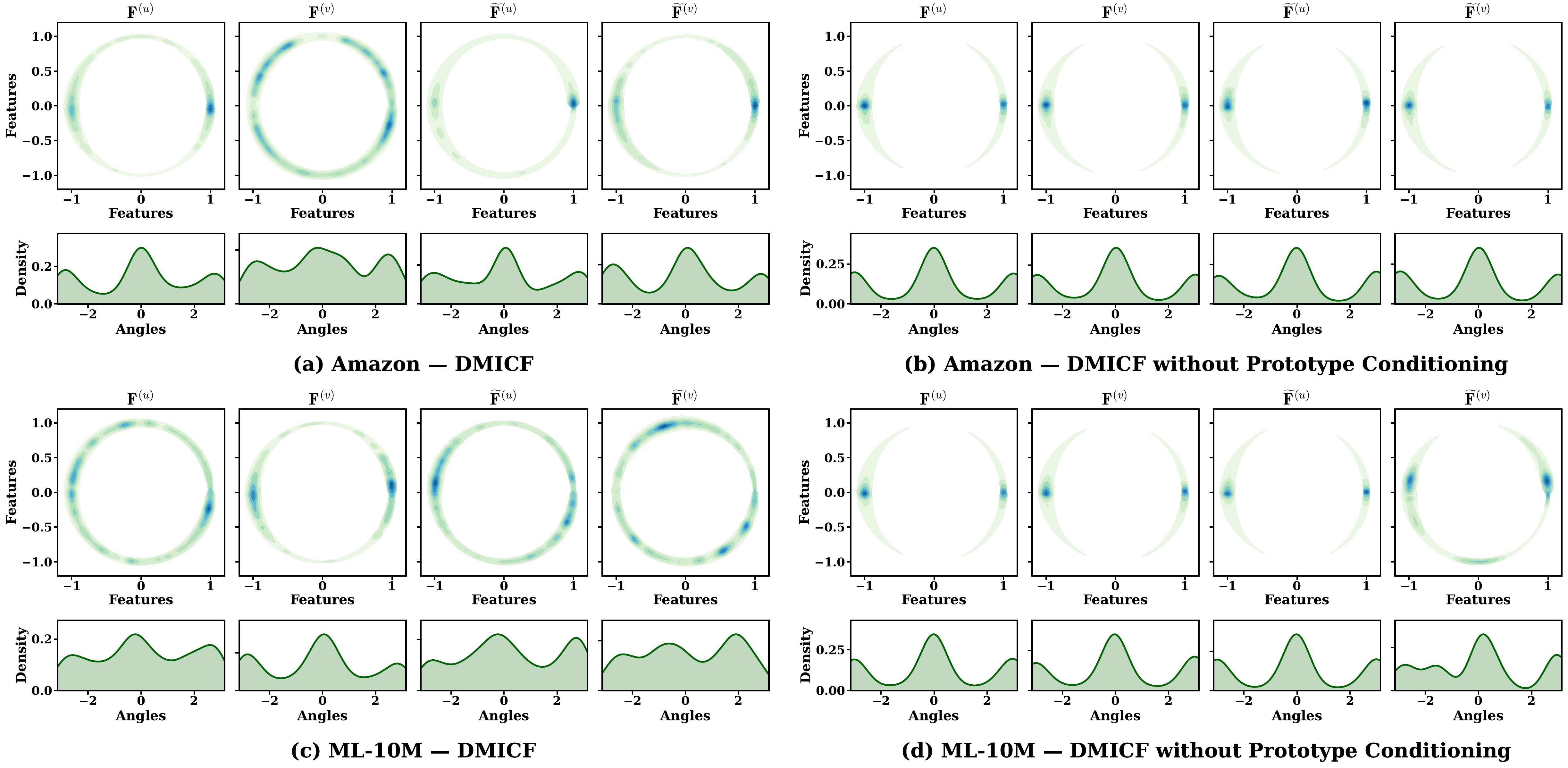}
  \caption{Latent intent embeddings $\textbf{F}^{(u)}$, $\textbf{F}^{(v)}$, $\widetilde{\textbf{F}}^{(u)}$, $\widetilde{\textbf{F}}^{(v)}$ generated by the variational encoder with/without prototype conditioning.}
  \label{fig:without_intent_prototypes}
\end{figure*}

\subsubsection{\textbf{Empirical Validation of Intent Anti-Collapse}}
\label{intent_anti_collapse}
Following \emph{SimGCL}~\cite{SimGCL}, we sample 2,000 user/item representations and visualize their spatial and angular density distributions in Fig.~\ref{fig:without_intent_prototypes}. We provide the implementation code for reproducibility. The results show that prototype-aware variational conditioning effectively mitigates intent collapse. With semantic prototype conditioning, \emph{DMICF} forms multiple well-separated high-density regions on the unit hypersphere, accompanied by clearly multi-peaked angular distributions. This indicates that the aggregate posterior is organized into localized semantic modes rather than collapsing into a single region, consistent with \textbf{Proposition~\ref{proposition}}, which characterizes it as an implicit prototype-anchored mixture. In contrast, removing prototype conditioning ($w/o\,IP$) leads to highly concentrated spatial distributions and near-unimodal angular densities, a characteristic manifestation of posterior collapse. By preserving both inter-mode separation and smooth intra-mode variation, \emph{DMICF} avoids semantic homogenization while retaining personalized intent diversity.

\section{RELATED WORK}

\textbf{CF-Based Recommendation Models.} 
Graph-based CF has evolved from early GCN-based models to more efficient architectures and self-supervised learning (SSL)-enhanced variants. Early models such as \emph{NGCF}~\cite{NGCF} propagate high-order connectivity via multi-hop message passing, while \emph{GCCF}~\cite{GCCF} and \emph{LightGCN}~\cite{LightGCN} simplify propagation by removing nonlinearities and transformations. However, these approaches struggle under sparse interactions due to limited structural signals. To alleviate sparsity, subsequent works exploit higher-order homophily through user–user or item–item relations (e.g., \emph{Multi-GCCF}~\cite{Multi-GCCF}, \emph{AdaMCL}~\cite{AdaMCL}). SSL has also been introduced to improve representation robustness, including contrastive learning (\emph{LightGCL}~\cite{LightGCL}, \emph{SimGCL}~\cite{SimGCL}, \emph{RGCL}~\cite{RGCL}, \emph{SCCF}~\cite{SCCF}, \emph{LightCCF}~\cite{LightCCF}) as well as alternative objectives such as knowledge distillation (\emph{SimRec}~\cite{SimRec}) or mutual information maximization (\emph{STERLING}~\cite{STERLING}). Perturbation-based regularization methods (AU$^{+}$~\cite{AUPlus} and \emph{PAAC}~\cite{PAAC}) mitigate overfitting and popularity bias. While effective, these methods focus primarily on structural refinement. In practice, user behaviors are driven by multiple latent and entangled intents. Without explicit disentanglement, representations conflate heterogeneous interaction signs, limiting personalization and interpretability. This motivates intent-aware models that disentangle and align latent semantics beyond structural proximity.

\textbf{Disentanglement-based Recommendation.} 
Disentangled recommenders aim to decompose user/item representations into latent factors, each capturing a distinct semantic aspect of interactions. Representative approaches include autoencoding-based models (e.g., \emph{MacridVAE}~\cite{MacridVAE}, \emph{DualVAE}~\cite{DualVAE}, and \emph{VARIUM}~\cite{varium}), graph-based methods (\emph{DGCF}~\cite{DGCF} and \emph{DisenHAN}~\cite{DisenHAN}), as well as extensions incorporating local–global signals or heterogeneous attention mechanisms (\emph{MIDGN}~\cite{MIDGN}, \emph{LGD-GCN}~\cite{LGDGCN}). Knowledge-aware frameworks (\emph{KGIN}~\cite{KGIN}, \emph{KDR}~\cite{KDR}, \emph{DIKGNN}~\cite{DIKGNN}, and \emph{IntentPCA}~\cite{IntentPCA}) leverage external graphs for semantic supervision, while geometric or semantic priors are explored in \emph{GDCF}~\cite{GDCF}, \emph{UDITSR}~\cite{UDITSR}, and \emph{GDCCDR}~\cite{GDCCDR}. Recently, contrastive learning has further been adopted to enhance robustness, with methods such as \emph{DCCF}, \emph{BIGCF}, \emph{IHGCL}~\cite{IHGCL}, and \emph{IPCCF} aligning node representations under different structural perturbations or propagation schemes. However, most existing models fail to preserve perspective-dependent interaction semantics due to single-space interaction modeling. Moreover, user–item intent alignment lacks explicit interaction-level supervision and is prone to semantic drift during interaction instantiation. These limitations motivate \emph{DMICF}, which combines dual-perspective structural encoding with interaction-driven intent alignment to enable fine-grained, perspective-consistent disentanglement.

\section{CONCLUSION}
We proposed \emph{DMICF}, a dual-perspective variational framework for disentangled intent modeling in collaborative filtering. By integrating macro–micro prototype-aware encoding with interaction-level intent alignment, \emph{DMICF} learns fine-grained and uncertainty-aware intent representations. Experiments show consistent improvements over strong baselines. Ablation studies indicate robustness to variations in individual module designs, provided that dual-perspective modeling and interaction-level intent alignment are preserved.


\bibliographystyle{ACM-Reference-Format}
\bibliography{sample-base}

@String{Computing = "Computing" }

@article{IPCCF,
 author = {Li, Haojie and Du, Junwei and Liu, Guanfeng and Jiang, Feng and Wang, Yan and Zhou, Xiaofang},
 title = {Intent Propagation Contrastive Collaborative Filtering},
 journal = {IEEE Transactions on Knowledge and Data Engineering},
 volume = {37},
 number = {5},
 year = {2025},
 pages = {2665--2679},
 doi = {10.1109/TKDE.2025.3543241},
}

@Inproceedings{BIGCF,
  author =       "Zhang, Yi and Sang, Lei and Zhang, Yiwen",
  title =        "Exploring the Individuality and Collectivity of Intents behind Interactions for Graph Collaborative Filtering",
  booktitle =    "Proceedings of the 47th International ACM SIGIR Conference on Research and Development in Information Retrieval",
  series =       "SIGIR '24",
  year =         2024,
  publisher =    "Association for Computing Machinery",
  address =      "New York, NY, USA",
  pages =        "1253--1262",
  doi =          "10.1145/3626772.3657738",
}

@Inproceedings{LightCCF,
  author =       "Zhang, Yu and Zhang, Yiwen and Zhang, Yi and Sang, Lei and Yang, Yun",
  title =        "Unveiling Contrastive Learning's Capability of Neighborhood Aggregation for Collaborative Filtering",
  booktitle =    "Proceedings of the 48th International ACM SIGIR Conference on Research and Development in Information Retrieval",
  series =       "SIGIR '25",
  year =         2025,
  publisher =    "Association for Computing Machinery",
  address =      "New York, NY, USA",
  pages =        "1985--1994",
  doi =          "10.1145/3726302.3730111",
}

@Inproceedings{DCCF,
  author =       "Ren, Xubin and Xia, Lianghao and Zhao, Jiashu and Yin, Dawei and Huang, Chao",
  title =        "Disentangled Contrastive Collaborative Filtering",
  booktitle =    "Proceedings of the 46th International ACM SIGIR Conference on Research and Development in Information Retrieval",
  series =       "SIGIR '23",
  year =         2023,
  publisher =    "Association for Computing Machinery",
  address =      "New York, NY, USA",
  pages =        "1137--1146",
  doi =          "10.1145/3539618.3591665",
}

@Inproceedings{IMP-GCN,
  author =       "Liu, Fan and Cheng, Zhiyong and Zhu, Lei and Gao, Zan and Nie, Liqiang",
  title =        "Interest-aware Message-Passing GCN for Recommendation",
  booktitle =    "Proceedings of the 46th International ACM SIGIR Conference on Research and Development in Information Retrieval",
  series =       "WWW '21",
  year =         2021,
  publisher =    "Association for Computing Machinery",
  address =      "New York, NY, USA",
  pages =        "1296--1305",
  doi =          "10.1145/3442381.3449986",
}

@article{SEM-VAE,
 author = {Wang, Xin and Chen, Hong and Zhou, Yuwei and Ma, Jianxin and Zhu, Wenwu},
 title = {Disentangled Representation Learning for Recommendation},
 journal = {IEEE Transactions on Pattern Analysis and Machine Intelligence},
 volume = {45},
 number = {1},
 year = {2023},
 pages = {408--424},
 doi = {10.1109/TPAMI.2022.3153112},
}

@Inproceedings{AdaMCL,
  author =       "Zhu, Guanghui and Lu, Wang and Yuan, Chunfeng and Huang, Yihua",
  title =        "AdaMCL: Adaptive Fusion Multi-View Contrastive Learning for Collaborative Filtering",
  booktitle =    "Proceedings of the 46th International ACM SIGIR Conference on Research and Development in Information Retrieval",
  series =       "SIGIR '23",
  year =         2023,
  publisher =    "Association for Computing Machinery",
  address =      "New York, NY, USA",
  pages =        "1076--1085",
  doi =          "10.1145/3539618.3591632",
}

@article{EGCF,
 author = {Zhang, Yi and Zhang, Yiwen and Sang, Lei and Sheng, Victor S.},
 title = {Simplify to the Limit! Embedding-Less Graph Collaborative Filtering for Recommender Systems},
 journal = {ACM Trans. Inf. Syst.},
 volume = {43},
 number = {1},
 year = {2024},
 pages = {1--30},
 doi = {10.1145/3701230},
}

@article{REC-SURVEY-1,
 author = {Ren, Xubin and Wei, Wei and Xia, Lianghao and Huang, Chao},
 title = {A Comprehensive Survey on Self-Supervised Learning for Recommendation},
 journal = {ACM Comput. Surv.},
 volume = {58},
 number = {1},
 year = {2025},
 pages = {1--38},
 doi = {10.1145/3746280},
}

@article{REC-SURVEY-2,
 author = {Zhang, Kaike and Cao, Qi and Sun, Fei and Wu, Yunfan and Tao, Shuchang and Shen, Huawei and Cheng, Xueqi},
 title = {Robust Recommender System: A Survey and Future Directions},
 journal = {ACM Comput. Surv.},
 volume = {58},
 number = {1},
 year = {2025},
 pages = {1--38},
 doi = {10.1145/3757057},
}

@article{REC-SURVEY-3,
 author = {Wu, Le and He, Xiangnan and Wang, Xiang and Zhang, Kun and Wang, Meng},
 title = {A Survey on Accuracy-Oriented Neural Recommendation: From Collaborative Filtering to Information-Rich Recommendation},
 journal = {IEEE Transactions on Knowledge and Data Engineering},
 volume = {35},
 number = {5},
 year = {2023},
 pages = {4425--4445},
 doi = {10.1109/TKDE.2022.3145690},
}

@article{CF-SURVEY,
 author = {G. Silva, Miguel and C. Madeira, Sara and Henriques, Rui},
 title = {A Comprehensive Survey on Biclustering-based Collaborative Filtering},
 journal = {ACM Comput. Surv.},
 volume = {56},
 number = {12},
 year = {2024},
 pages = {1--32},
 doi = {10.1145/3674723},
}

@Inproceedings{HCCF,
  author =       "Xia, Lianghao and Huang, Chao and Xu, Yong and Zhao, Jiashu and Yin, Dawei and Huang, Jimmy",
  title =        "Hypergraph Contrastive Collaborative Filtering",
  booktitle =    "Proceedings of the 45th International ACM SIGIR Conference on Research and Development in Information Retrieval",
  series =       "SIGIR '22",
  year =         2022,
  publisher =    "Association for Computing Machinery",
  address =      "New York, NY, USA",
  pages =        "70--79",
  doi =          "10.1145/3477495.3532058",
}

@Inproceedings{LightGCL,
  author =       "Cai, Xuheng and Huang, Chao and Xia, Lianghao and Ren, Xubin",
  title =        "LightGCL: Simple Yet Effective Graph Contrastive Learning for Recommendation",
  booktitle =    "The Eleventh International Conference on Learning Representations",
  year =         2023,
}

@Inproceedings{LightGCN,
  author =       "He, Xiangnan and Deng, Kuan and Wang, Xiang and Li, Yan and Zhang, YongDong and Wang, Meng",
  title =        "LightGCN: Simplifying and Powering Graph Convolution Network for Recommendation",
  booktitle =    "Proceedings of the 43rd International ACM SIGIR Conference on Research and Development in Information Retrieval",
  series =       "SIGIR '20",
  year =         2020,
  publisher =    "Association for Computing Machinery",
  address =      "New York, NY, USA",
  pages =        "639--648",
  doi =          "10.1145/3397271.3401063",
}

@Inproceedings{DirectAU,
  author =       "Wang, Chenyang and Yu, Yuanqing and Ma, Weizhi and Zhang, Min and Chen, Chong and Liu, Yiqun and Ma, Shaoping",
  title =        "Towards Representation Alignment and Uniformity in Collaborative Filtering",
  booktitle =    "Proceedings of the 28th ACM SIGKDD Conference on Knowledge Discovery and Data Mining",
  series =       "KDD '22",
  year =         2022,
  publisher =    "Association for Computing Machinery",
  address =      "New York, NY, USA",
  pages =        "1816--1825",
  doi =          "10.1145/3534678.3539253",
}

@Inproceedings{NGCF,
  author =       "Wang, Xiang and He, Xiangnan and Wang, Meng and Feng, Fuli and Chua, Tat-Seng",
  title =        "Neural Graph Collaborative Filtering",
  booktitle =    "Proceedings of the 42nd International ACM SIGIR Conference on Research and Development in Information Retrieval",
  series =       "SIGIR '19",
  year =         2019,
  publisher =    "Association for Computing Machinery",
  address =      "New York, NY, USA",
  pages =        "165--174",
  doi =          "10.1145/3331184.3331267",
}

@Inproceedings{DGCF,
  author =       "Wang, Xiang and Jin, Hongye and Zhang, An and He, Xiangnan and Xu, Tong and Chua, Tat-Seng",
  title =        "Disentangled Graph Collaborative Filtering",
  booktitle =    "Proceedings of the 43rd International ACM SIGIR Conference on Research and Development in Information Retrieval",
  series =       "SIGIR '20",
  year =         "2020",
  publisher =    "Association for Computing Machinery",
  address =      "New York, NY, USA",
  pages =        "1001--1010",
  doi =          "10.1145/3397271.3401137",
}

@Inproceedings{UDITSR,
  author =       "Zhang, Yuting and Wu, Yiqing and Han, Ruidong and Sun, Ying and Zhu, Yongchun and Li, Xiang and Lin, Wei and Zhuang, Fuzhen and An, Zhulin and Xu, Yongjun",
  title =        "Unified Dual-Intent Translation for Joint Modeling of Search and Recommendation",
  booktitle =    "Proceedings of the 30th ACM SIGKDD Conference on Knowledge Discovery and Data Mining",
  series =       "KDD '24",
  year =         "2024",
  publisher =    "Association for Computing Machinery",
  address =      "New York, NY, USA",
  pages =        "6291--6300",
  doi =          "10.1145/3637528.3671519",
}

@Inproceedings{DisenHAN,
  author =       "Wang, Yifan and Tang, Suyao and Lei, Yuntong and Song, Weiping and Wang, Sheng and Zhang, Ming",
  title =        "DisenHAN: Disentangled Heterogeneous Graph Attention Network for Recommendation",
  booktitle =    "Proceedings of the 29th ACM International Conference on Information \& Knowledge Management",
  series =       "CIKM '20",
  year =         "2020",
  publisher =    "Association for Computing Machinery",
  address =      "New York, NY, USA",
  pages =        "1605--1614",
  doi =          "10.1145/3340531.3411996",
}

@Inproceedings{KGIN,
  author =       "Wang, Xiang and Huang, Tinglin and Wang, Dingxian and Yuan, Yancheng and Liu, Zhenguang and He, Xiangnan and Chua, Tat-Seng",
  title =        "Learning Intents behind Interactions with Knowledge Graph for Recommendation",
  booktitle =    "Proceedings of the Web Conference 2021",
  series =       "WWW '21",
  year =         "2021",
  publisher =    "Association for Computing Machinery",
  address =      "New York, NY, USA",
  pages =        "878--887",
  doi =          "10.1145/3442381.3450133",
}

@Inproceedings{GCCF,
  author =       "Chen, Lei and Wu, Le and Hong, Richang and Zhang, Kun and Wang, Meng",
  title =        "Revisiting Graph Based Collaborative Filtering: A Linear Residual Graph Convolutional Network Approach",
  booktitle =    "Proceedings of the AAAI Conference on Artificial Intelligence",
  year =         "2020",
  pages =        "27--34",
  doi =          "10.1609/aaai.v34i01.5330",
}

@Inproceedings{MacridVAE,
  author =       "Ma, Jianxin and Zhou, Chang and Cui, Peng and Yang, Hongxia and Zhu, Wenwu",
  title =        "Learning Disentangled Representations for Recommendation",
  booktitle =    "Proceedings of the 33rd International Conference on Neural Information Processing Systems",
  series =       "NIPS '19",
  year =         "2019",
  pages =        "5712--5723",
}

@Inproceedings{STERLING,
  author =       "Jing, Baoyu and Yan, Yuchen and Ding, Kaize and Park, Chanyoung and Zhu, Yada and Liu, Huan and Tong, Hanghang",
  title =        "STERLING: synergistic representation learning on bipartite graphs",
  booktitle =    "Proceedings of the AAAI Conference on Artificial Intelligence",
  year =         "2024",
  pages =        "12976--12984",
  doi =          "https://doi.org/10.1609/aaai.v38i12.29195",
}

@Inproceedings{AUPlus,
  author =       "Ouyang, Zhongyu and Zhang, Chunhui and Hou, Shifu and Zhang, Chuxu and Ye, Yanfang",
  title =        "How to Improve Representation Alignment and Uniformity in Graph-Based Collaborative Filtering?",
  booktitle =    "Proceedings of the International AAAI Conference on Web and Social Media",
  year =         "2024",
  pages =        "1148--1159",
  doi =          "10.1609/icwsm.v18i1.31379",
}

@Inproceedings{PAAC,
  author =       "Cai, Miaomiao and Chen, Lei and Wang, Yifan and Bai, Haoyue and Sun, Peijie and Wu, Le and Zhang, Min and Wang, Meng",
  title =        "Popularity-Aware Alignment and Contrast for Mitigating Popularity Bias",
  booktitle =    "Proceedings of the 30th ACM SIGKDD Conference on Knowledge Discovery and Data Mining",
  series =       "KDD '24",
  year =         "2024",
  publisher =    "Association for Computing Machinery",
  address =      "New York, NY, USA",
  pages =        "187--198",
  doi =          "10.1145/3637528.3671824",
}

@article{KDR,
 author = {Mu, Shanlei and Li, Yaliang and Zhao, Wayne Xin and Li, Siqing and Wen, Ji-Rong},
 title = {Knowledge-Guided Disentangled Representation Learning for Recommender Systems},
 journal = {ACM Trans. Inf. Syst.},
 volume = {40},
 number = {1},
 year = {2021},
 issn = {1046-8188},
 pages = {1--26},
 doi = {10.1145/3464304},
}

@Inproceedings{DIKGNN,
  author =       "Tu, Ke and Qu, Wei and Wu, Zhengwei and Zhang, Zhiqiang and Liu, Zhongyi and Zhao, Yiming and Wu, Le and Zhou, Jun and Zhang, Guannan",
  title =        "Disentangled Interest importance aware Knowledge Graph Neural Network for Fund Recommendation",
  booktitle =    "Proceedings of the 32nd ACM International Conference on Information and Knowledge Management",
  series =       "CIKM '23",
  year =         "2023",
  publisher =    "Association for Computing Machinery",
  address =      "New York, NY, USA",
  pages =        "2482--2491",
  doi =          "10.1145/3583780.3614846",
}

@Inproceedings{GDCCDR,
  author =       "Liu, Jing and Sun, Lele and Nie, Weizhi and Jing, Peiguang and Su, Yuting",
  title =        "Graph disentangled contrastive learning with personalized transfer for cross-domain recommendation",
  booktitle =    "Proceedings of the AAAI Conference on Artificial Intelligence",
  year =         "2024",
  pages =        "8769--8777",
  doi =          "10.1609/aaai.v38i8.28723",
}

@Inproceedings{SimGCL,
  author =       "Yu, Junliang and Yin, Hongzhi and Xia, Xin and Chen, Tong and Cui, Lizhen and Nguyen, Quoc Viet Hung",
  title =        "Are Graph Augmentations Necessary? Simple Graph Contrastive Learning for Recommendation",
  booktitle =    "Proceedings of the 45th International ACM SIGIR Conference on Research and Development in Information Retrieval",
  series =       "SIGIR '22",
  year =         "2022",
  pages =        "1294--1303",
  doi =          "10.1145/3477495.3531937",
}

@Inproceedings{RGCL,
  author =       "Shuai, Jie and Zhang, Kun and Wu, Le and Sun, Peijie and Hong, Richang and Wang, Meng and Li, Yong",
  title =        "A Review-aware Graph Contrastive Learning Framework for Recommendation",
  booktitle =    "Proceedings of the 45th International ACM SIGIR Conference on Research and Development in Information Retrieval",
  series =       "SIGIR '22",
  year =         "2022",
  pages =        "1283--1293",
  doi =          "10.1145/3477495.3531927",
}

@article{IntentPCA,
 author = {Ai, Yuyan and Li, Chaoqun and Jiang, Liangxiao},
 title = {Intent-aware Recommendation Based on Principal Component Analysis},
 journal = {ACM Trans. Knowl. Discov. Data},
 volume = {19},
 number = {5},
 year = {2025},
 issn = {1556-4681},
 pages = {1--21},
 doi = {10.1145/3731761},
}

@Inproceedings{MIDGN,
  author =       "Zhao, Sen and Wei, Wei and Zou, Ding and Mao, Xianling",
  title =        "Multi-view intent disentangle graph networks for bundle recommendation",
  booktitle =    "Proceedings of the 36th AAAI Conference on Artificial Intelligence",
  year =         "2022",
  pages =        "4379--4387",
  doi =          "10.1609/aaai.v36i4.20359",
}

@article{LGDGCN,
 author = {Guo, Jingwei and Huang, Kaizhu and Yi, Xinping and Zhang, Rui},
 title = {Learning Disentangled Graph Convolutional Networks Locally and Globally},
 journal = {IEEE Transactions on Neural Networks and Learning Systems},
 volume = {35},
 number = {3},
 year = {2024},
 pages = {3640--3651},
 doi = {10.1109/TNNLS.2022.3195336},
}

@article{IHGCL,
 author = {Sang, Lei and Wang, Yu and Zhang, Yi and Zhang, Yiwen and Wu, Xindong},
 title = {Intent-Guided Heterogeneous Graph Contrastive Learning for Recommendation},
 journal = {IEEE Transactions on Knowledge and Data Engineering},
 volume = {37},
 number = {4},
 year = {2025},
 pages = {1915--1929},
 doi = {10.1109/TKDE.2025.3536096},
}

@Inproceedings{SimRec,
  author =       "Xia, Lianghao and Huang, Chao and Shi, Jiao and Xu, Yong",
  title =        "Graph-less Collaborative Filtering",
  booktitle =    "Proceedings of the ACM Web Conference 2023",
  series =       "WWW '23",
  year =         "2023",
  pages =        "17--27",
  doi =          "10.1145/3543507.3583196",
}

@Inproceedings{GDCF,
  author =       "Zhang, Yiding and Li, Chaozhuo and Xie, Xing and Wang, Xiao and Shi, Chuan and Liu, Yuming and Sun, Hao and Zhang, Liangjie and Deng, Weiwei and Zhang, Qi",
  title =        "Geometric Disentangled Collaborative Filtering",
  booktitle =    "Proceedings of the 45th International ACM SIGIR Conference on Research and Development in Information Retrieval",
  series =       "SIGIR '22",
  year =         "2022",
  publisher =    "Association for Computing Machinery",
  address =      "New York, NY, USA",
  pages =        "80--90",
  doi =          "10.1145/3477495.3531982",
}

@Inproceedings{Multi-GCCF,
  author =       "Sun, Jianing and Zhang, Yingxue",
  title =        "Multi-Graph Convolutional Neural Networks for Representation Learning in Recommendation",
  booktitle =    "Proceedings of the 33rd International Conference on Neural Information Processing Systems",
  series =       "NIPS '19",
  year =         "2019",
}

@Inproceedings{varium,
  author =       "Tran, Nhu-Thuat and Lauw, Hady W.",
  title =        "VARIUM: Variational Autoencoder for Multi-Interest Representation with Inter-User Memory",
  booktitle =    "Proceedings of the Eighteenth ACM International Conference on Web Search and Data Mining",
  series =       "WSDM '25",
  year =         "2025",
  publisher =    "Association for Computing Machinery",
  address =      "New York, NY, USA",
  pages =        "187--198",
  doi =          "10.1145/3701551.3703558",
}

@Inproceedings{SCCF,
  author =       "Wu, Yihong and Zhang, Le and Mo, Fengran and Zhu, Tianyu and Ma, Weizhi and Nie, Jian-Yun",
  title =        "Unifying Graph Convolution and Contrastive Learning in Collaborative Filtering",
  booktitle =    "Proceedings of the 30th ACM SIGKDD Conference on Knowledge Discovery and Data Mining",
  series =       "KDD '24",
  year =         "2024",
  publisher =    "Association for Computing Machinery",
  address =      "New York, NY, USA",
  pages =        "3425--3436",
  doi =          "10.1145/3637528.3671840",
}

@Inproceedings{NLGCL,
  author =       "Xu, Jinfeng and Chen, Zheyu and Yang, Shuo and Li, Jinze and Wang, Hewei and Wang, Wei and Hu, Xiping and Ngai, Edith",
  title =        "NLGCL: Naturally Existing Neighbor Layers Graph Contrastive Learning for Recommendation",
  booktitle =    "Proceedings of the Nineteenth ACM Conference on Recommender Systems",
  series =       "RecSys '25",
  year =         "2025",
  publisher =    "Association for Computing Machinery",
  address =      "New York, NY, USA",
  pages =        "319--329",
  doi =          "10.1145/3705328.3748059",
}

@Inproceedings{item-base-cf,
  author =       "Sarwar, Badrul and Karypis, George and Konstan, Joseph and Riedl, John",
  title =        "Item-based collaborative filtering recommendation algorithms",
  booktitle =    "Proceedings of the Nineteenth ACM Conference on Recommender Systems",
  series =       "WWW '01",
  year =         "2001",
  publisher =    "Association for Computing Machinery",
  address =      "New York, NY, USA",
  pages =        "285--295",
  doi =          "10.1145/371920.372071",
}

@Inproceedings{user-base-cf,
  author =       "Breese, John S. and Heckerman, David and Kadie, Carl",
  title =        "Empirical analysis of predictive algorithms for collaborative filtering",
  booktitle =    "Proceedings of the Fourteenth Conference on Uncertainty in Artificial Intelligence",
  series =       "UAI '98",
  year =         "1998",
  publisher =    "Morgan Kaufmann Publishers Inc.",
  address =      "San Francisco, CA, USA",
  pages =        "43--52",
}

@Inproceedings{NR-GCF,
  author =       "Chen, Yijun and Li, Bohan and Li, Yicong and Song, Lixiang and Wang, Haofen and Wu, Wenlong and Zhuo, Junnan and Yin, Hongzhi",
  title =        "NR-GCF: Graph Collaborative Filtering with Improved Noise Resistance",
  booktitle =    "Proceedings of the 34th ACM International Conference on Information and Knowledge Management",
  series =       "CIKM '25",
  year =         "2025",
  publisher =    "Association for Computing Machinery",
  address =      "New York, NY, USA",
  pages =        "373--382",
  doi =          "10.1145/3746252.3761342",
}

@Inproceedings{ELBO,
  author =       "Liang, Dawen and Krishnan, Rahul G. and Hoffman, Matthew D. and Jebara, Tony",
  title =        "Variational Autoencoders for Collaborative Filtering",
  booktitle =    "Proceedings of the 2018 World Wide Web Conference",
  series =       "WWW '18",
  year =         "2018",
  publisher =    "International World Wide Web Conferences Steering Committee",
  address =      "Republic and Canton of Geneva, CHE",
  pages =        "689--698",
  doi =          "10.1145/3178876.3186150",
}

@Inproceedings{DualVAE,
  author =       "Guo, Zhiqiang and Li, Guohui and Li, Jianjun and Wang, Chaoyang and Shi, Si",
  title =        "DualVAE: Dual Disentangled Variational AutoEncoder for Recommendation",
  booktitle =    "Proceedings of the 2024 SIAM International Conference on Data Mining (SDM)",
  series =       "SDM '24",
  year =         "2024",
  pages =        "571--579",
  doi =          "10.1137/1.9781611978032.66",
}




\end{document}